\documentclass[journal=jpccck,manuscript=article]{achemso}

\pdfoutput=1
\usepackage[version=3]{mhchem}

\newcommand{\T}{\rule{0pt}{2.6ex}}
\newcommand{\B}{\rule[-1.2ex]{0pt}{0pt}}

\author{William R. French}
\affiliation[Vanderbilt University]{Department of Chemical and Biomolecular Engineering, Vanderbilt University, Nashville, TN}
\author{Christopher R. Iacovella}
\email{christopher.iacovella@vanderbilt.edu}
\affiliation[Vanderbilt University]{Department of Chemical and Biomolecular Engineering, Vanderbilt University, Nashville, TN}
\author{Peter T. Cummings}
\affiliation[Vanderbilt University]
{Department of Chemical and Biomolecular Engineering, Vanderbilt University, Nashville, TN}
\alsoaffiliation[Oak Ridge National Laboratory]{Center for Nanophase Materials Sciences, Oak Ridge National Laboratory, Oak Ridge, TN}

\title{The Influence of Molecular Adsorption on Elongating Gold Nanowires}

\begin{document}

\pagebreak

\begin{abstract}

Using molecular dynamics simulations, we study the impact of physisorbing adsorbates on the structural and mechanical evolution of gold nanowires (AuNWs) undergoing elongation.  We used various adsorbate models in our simulations, with each model giving rise to a different surface coverage and mobility of the adsorbed phase.  We find that the local structure and mobility of the adsorbed phase remains relatively uniform across all segments of an elongating AuNW, except for the thinning region of the wire where the high mobility of Au atoms disrupts the monolayer structure, giving rise to higher solvent mobility.  We analyzed the AuNW trajectories by measuring the ductile elongation of the wires and detecting the presence of characteristic structural motifs that appeared during elongation.  Our findings indicate that adsorbates facilitate the formation of high-energy structural motifs and lead to significantly higher ductile elongations.  In particular, our simulations result in a large number of monatomic chains and helical structures possessing mechanical stability in excess of what we observe in vacuum.  Conversely, we find that a molecular species that interacts weakly (i.e., does not adsorb) with AuNWs worsens the mechanical stability of monatomic chains.  

\end{abstract}

\textsc{}

\vspace{1in}

Keywords: Molecular Break Junctions, Molecular Electronics, Molecular Dynamics Simulations, Mechanically Controllable Break Junctions, Single-Molecule Measurements, Electron Transport.

\pagebreak

\section{Introduction}

Understanding the structures of 0-d and 1-d nanomaterials is an important step towards designing and tailoring the properties of materials on the nanoscale \cite{Pan:2010}.  The gold nanowire (AuNW) is an example of a 1-d nanomaterial whose structures and properties have been widely studied \cite{Rubio:1996,Scheer:1998,Yanson:1998,Marszalek:2000,Tsutsui_nl:2008,Rodrigues:2001,Rodrigues:2000,Oshima:2003,Kondo:2000,Kondo:1997,He:2002,Huo:2008,Coura:2004,Pu:2008,Pu_JCP:2007,Sato:2005,Lin:2005,Wang:2007,Koh:2006,Wang:2001,Wen:2010,Weinberger:2010} and utilized for novel nanoscale applications (e.g., nanowelding \cite{Lu:2010}, nanosprings \cite{Xu_nanosprings:2010}, and molecular sensors \cite{Shi:2010}).  In particular, the electrical and mechanical behaviors of AuNWs have garnered considerable attention due to unique structure-dependent features that are observed during elongation of the wires.  Measurements of conductance and force of an elongating AuNW exhibit stepwise (i.e., quantized) \cite{Rubio:1996,Scheer:1998,Yanson:1998,Marszalek:2000,Tsutsui_nl:2008} and sawtooth-like \cite{Marszalek:2000} changes, respectively, suggesting that the structure of AuNWs undergoes sudden atomic rearrangements as a mechanism for relieving the strain induced through stretching. Experimentally, various high-energy structures have been observed during the stretching process, including disordered chunks \cite{Kondo:1997}, helical ribbon-like structures \cite{Coura:2004,Oshima:2003,Kondo:2000,Rodrigues:2000}, and monatomic chains \cite{Coura:2004,Rodrigues:2001,Rodrigues:2000}.  A number of strategies have been employed in an attempt to facilitate the formation of these characteristic high-energy structures.  Rodrigues \emph{et al.} \cite{Rodrigues:2000} observed remarkable reproducibility of the appearance of monatomic chains when AuNWs were elongated in the [100] or [111] directions.  Pu and co-workers \cite{Pu:2008} used molecular dynamics simulations to show that the emergence of long monatomic chains is promoted by elongation rates that are high enough to preclude the relaxation of low-energy defect modes in the system.  

Observing important structural events at the atomic level, as well as precisely controlling the conditions under which a AuNW is stretched, can be extremely challenging in an experimental setting.  For example, despite important experimental progress that now enables electron transport measurements to be made on the nanosecond timescale \cite{Guo:2011}, such measurements remain too slow for observing the breakdown event of a AuNW.  As an alternative to running experiments, atomic-based simulations can be performed to study AuNW elongation.  In simulations the features of a system can be controlled and measured with femtosecond and sub-angstrom resolution.  A vast majority of simulation studies to date have focused on AuNWs undergoing elongation in vacuum \cite{Coura:2004,Pu:2008,Pu_JCP:2007,Sato:2005,Lin:2005,Wang:2007,Koh:2006,Wang:2001,Wen:2010}, even though many important applications are conducted in non-vacuum environments.  One notable example is metal-molecule-metal mechanically controllable break junction (MCBJ) experiments \cite{Reed:1997,Wu:2008,Tsutsui_apl:2008,Tsutsui_nl:2009,Schneebeli:2011}, in which a AuNW undergoes chemisorption with an adsorbate prior to elongation.  By performing simulations of AuNW elongation in an adsorbate, realistic geometries for subsequent first-principles calculations of conductance can be obtained.  Direct force calculations have already played a large role in an ongoing effort to reconcile experimental measurements with theoretical predictions of the conductance through metal-molecule-metal structures \cite{Andrews:2008,Paulsson:2009,Strange:2010,Velez:2010,Kim:2010,Sergueev:2010}.

Moreover, the presence of an adsorbate may lead to the appearance of certain structural motifs preferentially over others in elongating AuNWs.  He \emph{et al.} \cite{He:2002} performed scanning tunneling microscopy-based experiments to test the effect of molecular adsorption on monatomic chain stability, finding that high-energy structures are stabilized by the presence of a strong adsorbate.  In these experiments, the presence and stability of monatomic chains were inferred from conductance measurements through the AuNWs.  However, previous work in the same group showed that the presence of adsorbate molecules can reduce the conductance through monatomic chains by as much as 50\% \cite{Bogozi:2001}, making the results of He \emph{et al.} somewhat ambiguous.  Thus, validating the conclusions of He \emph{et al.} by directly observing the structural features of elongating AuNWs in adsorbates would be valuable.

To this end, we employ atomic and molecular-level simulations to investigate the effect of adsorbates on the structural and mechanical behavior of elongating AuNWs.  We limit our study to adsorbates that physisorb to AuNWs, testing both a simple Lennard-Jones sphere model and an atomistically resolved model of propane.  In both cases, we simulate several levels of interaction strength between the adsorbate and Au atoms, enabling us to systematically assess the effect of a broad range of adsorbates.  To gather a comprehensive and statistical view of the effect of molecular adsorption on AuNW elongation, we carry out a large number of simulations (390 in total) spanning different wire sizes and adsorbate-adsorbent interaction strengths.  The results of this study allow for direct comparisons between various conditions that are applicable to real systems.

\section{Methodology}

The development of a physically accurate and robust model is the first step towards developing a design tool for systems of interest in nanoscience. Previous work in our group focused on developing, testing, and validating models and methods for AuNW elongation in vacuum \cite{Pu_JCP:2007} and in solvent \cite{Pu:2007}.  We apply and build upon these methods in this paper, as described below.

\subsection{Adsorbate Models}

Gold surfaces are known to interact strongly with several functional groups (e.g., thiols, amines, and carboxylic acids) \cite{Chen:2006}. Thiols (i.e., -SH functional groups) in particular are widely used in the formation of nanoscale structures due to their ability to form strong linkages with metallic surfaces \cite{Pontes:2006}.  Beyond such well-known anchoring groups, various other functional groups (e.g., sp$^2$ carbon and sp$^2$ nitrogen) \cite{Bilic:2006,Cafe:2007} exhibit strong physical interactions and weak chemical interactions with Au surfaces.  For example, Schneebeli \emph{et al.} \cite{Schneebeli:2011} recently performed single-molecule conductance measurements on ``anchor-less'' molecules by taking advantage of the strong interactions between strained benzene rings and a Au break junction.     

In this paper the selection of adsorbates was limited to those that physisorb to Au surfaces.  Explicit modeling of a chemisorbed species requires a more sophisticated simulation methodology \cite{Pu:2010} in order to accommodate the interface between the metallic surface and reactive headgroup.  A recent study based on post-Hartree-Fock calculations found that the dominant interaction between a Au (111) surface and a physical adsorbate was dispersion \cite{Piana:2006}.  To model this interaction, we used the 12-6 Lennard-Jones (LJ) potential:  

\begin{equation}U_{LJ}(r) = 4\epsilon_{ij}[(\frac{\sigma_{ij}}{r})^{12}-(\frac{\sigma_{ij}}{r})^{6}]\label{ULJ}
\end{equation} 

\noindent where $\epsilon_{ij}$ is the potential well depth and $\sigma_{ij}$ is the inter-atomic distance at which the energy between atoms $i$ and $j$ is zero.  The Au atoms were treated as uncharged particles and the effects of Au polarizability, which should be minimal in the presence of a non-polar adsorbate, were neglected. 

Two models were employed to simulate an adsorbate species.  The first model was a LJ sphere model, in which the adsorbate was represented as a single Lennard-Jones sphere (see Figure 1a).  The naming convention LJ/Prop-X is used for the LJ sphere model, where X is the ratio of the adsorbate-Au LJ well depth, $\epsilon_{Ads.-Au}$,  to the adsorbate-adsorbate well depth, $\epsilon_{Ads.-Ads.}$.  Thus, LJ/Prop-2.0 refers to a system in which $\epsilon_{Ads.-Au}$ exceeds $\epsilon_{Ads.-Ads.}$ by a factor of two. In accordance with our previous study \cite{Pu:2007}, the adsorbate-adsorbate LJ parameters ($\sigma$ = 4.66 \AA, $\epsilon$ = 0.553 kcal/mol) were fit to the critical parameters of propane.  The adsorbate-adsorbate parameters were held constant for all LJ/Prop-X simulations.  Propane was chosen because of its simple structure and because it exists as a liquid at room temperature and sufficiently high pressures. While the LJ parameters of the LJ/Prop-X model were chosen to reproduce the properties of liquid propane, by adjusting $\epsilon_{Ads.-Au}$, as reported in this work, a more general case adsorbate was modeled. The values of $\epsilon_{Ads.-Au}$/$\epsilon_{Ads.-Ads.}$ used in AuNW elongation simulations are listed in Table 1 and represent realistic interaction strengths for species that exhibit strong physical interactions with Au surfaces (e.g., sp$^2$ carbon and sp$^2$ nitrogen).  The manual adjustment of $\epsilon_{Ads.-Au}$ can also be viewed as analogous to inducing adsorption by altering the electrochemical potential of AuNWs in real experiments \cite{He:2002}.  Systems in which $\epsilon_{Ads.-Au}$/$\epsilon_{Ads.-Ads.}$ < 1.0 represented cases where the propane molecules preferred to interact with one another rather than the AuNW.  Such molecules can be thought of as a generic solvent that collides with (instead of adsorbing to) the AuNW during the elongation process.  This was the scenario studied in our previous work \cite{Pu:2007}.

Next, a more detailed adsorbate model was studied: all-atom propane (see Figure 1b).  The naming convention AA/Prop-Y is used for the all-atom model, where Y represents the study in literature from which the C-Au and H-Au LJ parameters were taken.   The AA/Prop-UFF and AA/Prop-FCC parameters were obtained by mixing C and H parameters from the OPLS-AA force field with Au parameters from Refs.~\cite{Rappe:1992,Heinz:2008} using geometric mixing rules.  The AA/Prop-MP2 C-Au and H-Au parameters were calculated explicitly in Ref.~\cite{Piana:2006} using Moller-Plesset second-order perturbation theory (MP2).  The use of these three separate sets of parameters allowed varying levels of interaction strength between propane and gold to be simulated (see Table 1).  The adsorbate-adsorbate interaction parameters were held fixed for all AA/Prop-Y simulations.  Partial charges, intermolecular parameters, and intramolecular parameters were taken from the OPLS-AA force field \cite{Jorgensen:1996}.  To estimate the adsorbate-Au well depth, $\epsilon_{Ads.-Au}$, of the all-atom models, the energy between an isolated adsorbate molecule and a single Au atom was computed at various fixed distances; the same procedure was followed for two adsorbate molecules to estimate $\epsilon_{Ads.-Ads.}$.  Results are listed in Table 2.

\subsection{Au-Au Interactions}
Simple pairwise potentials fail to capture many important properties (e.g., elastic constants, vacancy formation energies, surface structure, and relaxation properties) of metallic systems \cite{Leach:2001,Cleri:1993}.  These deficiencies are especially problematic for systems undergoing mechanical deformation.  For such systems, many-body potentials that approximate the local electron density are better suited for describing the band character of the metallic elements. We used the second-moment approximation to the tight-binding potential (TB-SMA), a common many-body semi-empirical potential, to compute the Au-Au interactions \cite{Cleri:1993}.  TB-SMA contains a band term that relates the electronic structure to the atomic positions, and has been shown to reproduce the energetic and structural features of AuNWs undergoing elongation better than other popular semi-empirical potentials, when compared to Density Functional Theory calculations \cite{Pu_JCP:2007}.  Furthermore, TB-SMA is a rather efficient many-body potential, only $\sim$4 times as costly as a simple LJ potential.  

\subsection{Simulation Details}

As in our previous work \cite{Pu_JCP:2007,Pu:2007,Pu:2008}, the continuous stretching of a AuNW was approximated using a stretch-and-relax molecular dynamics (MD) technique in which two layers of rigid ``grip'' atoms on one end of the wire were periodically displaced a small amount (0.1 \AA) in the [100] direction.  Two additional layers of rigid ``grip'' atoms were placed on the opposite end of the wire, while atoms within the core of the wire were dynamic.  The core atoms were allowed to relax for 10 ps between displacements of the ``grip'' layer atoms, corresponding to a nominal elongation rate of 1.0 m/s.  

Three [100]-oriented AuNWs with different initial diameters were used for the simulations.  Each of the AuNWs was initially 12.3 nm in length, which is long enough to avoid boundary effects.  The AuNWs were generated by taking cylindrical cuts from a FCC lattice of diameter 1.1, 1.5, and 1.9 nm, as shown in Figure 2.  While in this paper we have chosen to focus on diameter-dependent behavior for wires oriented in the [100] direction, AuNWs oriented in the [110] and [111] directions are also commonly used in experimental studies \cite{Coura:2004}.  Such wires are known to exhibit slightly different mechanical and structural behavior when elongated, and an adsorbed monolayer would likely have a slightly different structure and mobility on the surfaces of such wires.  The effect of adsorbates on wires with different crystallographic orientations will be addressed in a future publication.

To prepare the simulation boxes, isobaric-isothermal (constant $NPT$) MD simulations of pure propane were first performed at a target temperature of 298 K and pressure of 42.5 bar.  These conditions were chosen in order to obtain a dense liquid state of propane.  The boxes of propane were sufficiently long in the direction of pulling to allow for elongation and rupture of the wire.  Once the box of propane was equilibrated at constant temperature and pressure, the AuNW was inserted into the center of the box. Propane molecules overlapping with the wire were then removed from the box.  Before stretching the wire, the propane and AuNW were allowed to equilibrate in the canonical ensemble (constant $NVT$) for 50 ps.  Elongation of the AuNWs was also performed in the canonical ensemble.       

Simulations were performed within the LAMMPS simulation package \cite{Plimpton:1995}, with periodic boundary conditions applied in all three directions and temperature controlled through application of the Nos\'{e}-Hoover thermostat.  A non-bonded cutoff distance of 10 \AA\ was utilized for inter-particle computations.  While electrostatics were not included in the LJ/Prop-X systems, a Coulombic potential was used to compute the short-range electrostatic interaction between adsorbate molecules in the AA/Prop-Y systems.  Long-range electrostatic corrections were computed with the particle-particle particle-mesh solver (precision of 1.0 x 10$^{-5}$).  For the LJ/Prop-X systems, the equations of motion were integrated via the velocity Verlet algorithm with a time step of 1.0 fs.  For the AA/Prop-Y systems, the equations of motion were integrated via the velocity Verlet algorithm combined with the rRESPA multi-timescale integrator, with an outer loop time step of 1.0 fs (for intermolecular computations) and an inner loop time step of 0.2 fs (for intramolecular computations).  Excellent energy conservation was obtained for this integration scheme.

\section{Results and Discussion}

\subsection{Characterization of Adsorbed Phases on Unstretched Wires}

Altering the strength of interaction between an adsorbate and AuNW gives rise to different adsorbed phase behaviors.  As one might expect, stronger interactions result in a fluid that is more tightly bound to the surface.  This can be seen in Figure 3, where the top images depict the monolayers on a 1.1nm AuNW resulting from application of the AA/Prop-MP2 and AA/Prop-FCC models.  The bottom images in Figure 3 depict the molecules that 200 ps prior were adsorbed to the AuNW surface.  While the molecules in the AA/Prop-MP2 system predominantly detach from the AuNW surface, a vast majority of the molecules in the AA/Prop-FCC system remain on the surface after 200 ps.  A stronger interaction leads to higher surface densities, which, in turn, results in a lower mobility of the adsorbate on the AuNW surface while also suppressing the flux of molecules into and out of the monolayer.

Quantifying such differences can provide a convenient means of selecting or designing an adsorbate to yield desired AuNW behavior, as it is likely that the properties of the adsorbed phase correlates with the structural or mechanical behavior of elongating AuNWs.  MD simulations of propane molecules around a fixed AuNW were performed to calculate the equilibrium properties of the monolayer.  Following 150 ps of equilibration, trajectories were saved every 10 fs for 200 ps to calculate various properties of the adsorbed phase.

\subsubsection{Structural Behavior of Monolayer}

First, the structure of the adsorbate was analyzed by computing the distance of the adsorbate from the AuNW surface.  The distance from the surface of each adsorbate molecule was computed by taking the average of the pair distances between the adsorbate center of mass and the three nearest Au atoms.  Representative curves for three different LJ/Prop-X systems are shown in Figure 4, with the bulk density for the LJ/Prop-X model also plotted for reference.  The bulk density for the LJ/Prop-X model is 457 kg/m$^3$, which compares well to previous MD results of liquid propane run at slightly different conditions \cite{Krishna:2005}.  For all three systems, a sharp initial peak (corresponding to the adsorbed monolayer) rises well above the bulk density value at a distance from the surface of 4.22 \AA.  The initial peak is higher for systems in which the adsorbate-Au interaction is stronger.  The second peak is shifted towards the AuNW surface for stronger adsorbate-Au interactions.  The trough that occurs between the first and second peaks is similarly shifted.  This trough occurs at a distance from the nanowire, $r_\mathrm{boundary}$, that is taken to be the monolayer boundary and varies from system to system.  The narrowing of the first peak for higher interaction strengths signifies that the monolayer is packed more tightly in the radial direction around the AuNW.  A slight drop in density below the bulk value occurs at the box boundaries due to the adsorption of molecules onto the AuNW and out of the bulk phase.

The surface coverage, $\Theta$, was next computed by dividing the number of molecules in the monolayer (i.e., molecules located at a distance from the surface of less than $r_\mathrm{boundary}$) by the number of Au surface atoms.  Surface coverage is dictated by the interplay between entropic (packing and conformational) and enthalpic (adsorbate-adsorbent and adsorbate-adsorbate interactions) factors \cite{Jayaraman:2008,Jayaraman_Langmuir:2008}.  For calculations of $\Theta$, molecules at the ends of the wire were not considered to be apart of the monolayer, so the reported values represent coverages for a AuNW that extends infinitely in the [100] direction.  Results for three different wire sizes are shown in Figure 5.  The surface coverage data is fit to a Langmuir isotherm-type functional form: 

\begin{equation} \Theta = \alpha \frac{X}{1+X} + \beta \label{langmuir}\end{equation} where $X$ = $\epsilon_{Ads.-Au}$/$\epsilon_{Ads.-Ads.}$ and $\alpha$ and $\beta$ are fitting parameters.  Physically, $\alpha$ may be thought of as the gain in surface coverage obtained through increases in the adsorbate-Au interaction strength, while $\beta$ represents the surface coverage at vanishingly small adsorbate-Au interaction strengths.  Previous experimental measurements have reported coverages between 0.18-0.33 for physisorbed and chemisorbed monolayers on Au(111) surfaces \cite{Dubois:1992,Cafe:2007}.  For Au nanoparticles with core diameters of less than 2 nm (representing a highly curved surface), values for $\Theta$ between 0.68-0.80 have been reported both experimentally \cite{Hostetler:1998} and theoretically \cite{Jimenez:2010} for chemisorbed monolayers.  Our values for $\Theta$ on a 1.1nm AuNWs at high adsorbate-Au interaction strengths agree with these previously reported data.  Results for the larger AuNWs are lower than previously reported data.  Further increases in interaction strengths (e.g., up to typical values for chemisorption) would likely yield better agreement.  At low interaction strengths the propane does not adsorb to the AuNW, thus a low surface coverage is obtained.  Molecules close to the AuNW are likely to be pulled away from the surface if $\epsilon_{Ads.-Au}$ < $\epsilon_{Ads.-Ads.}$.  These molecules can be thought of as a generic solvent species that collides with AuNWs during the elongation process.  Additionally, thermal-induced desorption may occur when $\epsilon_{Ads.-Au}$/$\epsilon_{Ads.-Ads.}$ is less than or comparable to k$_b$T/$\epsilon_{Ads.-Ads.}$ = 1.07.

It is well-known that increasing curvature of a surface enables higher monolayer coverages due to increases in accessible free volume of the ligand tailgroups \cite{Hostetler:1998,Singh:2007,Jimenez:2010}.  Our results show that this trend holds for adsorbates on AuNWs.  Not only do smaller wires enable higher coverages, they also yield larger gains in surface coverage with increases in the adsorbate-adsorbent interaction strength.  This behavior is quantified by the $\alpha$ parameter in equation 2.  Table 3 shows values for the fitting parameters obtained for the LJ/Prop-X model on different AuNW sizes.  $\alpha$ is significantly higher for the two smaller wires than the larger one.  This large jump indicates that the larger [100] faces on the 1.9nm wire surface enable increased intra-monolayer interactions. 

\subsubsection{Mobility of Monolayer}

In addition to measuring the structure of the adsorbed phase, the mobility of the adsorbate molecules was also analyzed.  The mobility of the adsorbed phase was probed by calculating two properties: ($i$) adsorption rates and ($ii$) diffusion, $D_x$, across the AuNW surface.  Adsorption rates describe how molecules move radially around AuNWs while $D_x$ describes how adsorbate species move along the wire surface (in the [100] direction).

At equilibrium, the rate of adsorption of molecules onto a AuNW is equal to the rate of desorption from the AuNW surface.  While the surface coverage remains relatively constant, molecules in the monolayer are free to exchange with molecules in the bulk.  This rate of exchange (i.e., adsorption rate) was measured by counting the number of molecules that cross $r_\mathrm{boundary}$ during a 200 ps time frame.  The diffusion of molecules along the wire was measured using the Green-Kubo \cite{Frenkel:2002} relation:

\begin{equation}D_x = \int_{0}^{t_f} \! \Big\langle v_x(t)v_x(0) \Big\rangle \, dt\label{diffusion}\end{equation} where $v_x$ and $D_x$ are the velocity and average diffusion coefficient in the [100] direction, respectively, of an individual monolayer molecule.  While a value approaching infinity for $t_f$ is ideal for bulk diffusion calculations, a relatively small value of $t_f$ is appropriate for calculating properties of a monolayer.  A smaller value is appropriate because diffusion is a function of the local density around a AuNW and a molecule adsorbed to a AuNW at one instant time can desorb only a short time later.  A value of $t_f$ = 2 ps was selected and the 100 molecules that were on average closest to the AuNW during each successive 2 ps time span were used to compute the total average for $D_x$.  

As shown in Figure 6, the adsorption rate and $D_x$ both decrease as the adsorbate-Au interaction strength is increased.  The curves in Figure 6 are both exponential fits, indicating that both adsorption rate and $D_x$ behave as $\sim$$e^{\alpha X}$, where $\alpha$ is a constant and X = $\epsilon_{Ads.-Au}/\epsilon_{Ads.-Ads.}$.  This is consistent with Arrhenius behavior, which would predict mobility of the form $\sim$$e^{A/kT}$, where $A$ is the activation energy, $k$ is the Boltzmann constant, and $T$ is the temperature. The AA/Prop-Y model results fall mostly below the LJ/Prop-X model results for both measures of mobility.  This is likely due to the AA/Prop-Y molecules needing to orient themselves properly in order to diffuse towards and along the AuNW, whereas the LJ/Prop-X molecules experience no such orientation-dependence.  Figure 6 also shows $D_x$ plotted as a function of adsorbate-Au interaction strength.  The measured bulk diffusion value for the LJ/Prop-X model is 1.55 x 10$^{-8}$ m$^2$/s, which is slightly higher than the value of 1.05 x 10$^{-8}$ m$^2$/s reported elsewhere for MD simulations of liquid propane \cite{Krishna:2005}.  This discrepancy can be attributed to differences in the conditions of the two simulations, which result in a higher propane density (548 kg/m$^3$) in Ref. \cite{Krishna:2005} than the value we obtain (457 kg/m$^3$).

\subsection{Adsorbate Behavior on Elongating Wires}

Adsorbates are likely to respond differently to the presence of different facets and non-crystalline domains that emerge on the wire surface during elongation.  To investigate this, we measure the structure and mobility as a function of the local wire structure during elongation.  

\subsubsection{Surface Coverage and Desorption Residence Time}

Results for a 1.9nm AuNW elongated in the presence of LJ/Prop-2.0 adsorbate are presented in Figure 7.  Properties of the monolayer are measured and presented at different stages of elongation.  In Figure 7 (left) the density (relative to the bulk density) of LJ/Prop-2.0 around a AuNW is presented as color intensity maps.  Adsorbate density was measured along the 2.3nm segment enclosed by the solid lines at the bottom of each image in Figure 7 (right).  This segment corresponds the thinning region of the wire at 30 {\AA} of elongation.  For consistency, data were averaged along this same segment for each stage of elongation.  While calculating adsorption rates onto unstretched wires is straightforward, calculating adsorption rates onto elongating AuNWs is challenging since this requires an approximation of the non-uniform area surrounding the wire through which molecules move in and out of the monolayer.  Therefore, the mobility of the adsorbed phase was instead quantified by calculating the residence time correlation function \cite{Yang:2010}:  

\begin{equation}R(t)= \Bigg\langle \frac{1}{N}\sum_{i=1}^{N}\theta(t_{0})\theta(t_{0}+t) \Bigg\rangle \,\label{residence_correlation}\end{equation} where $\theta$($t_{0}$) equals one when a monolayer molecule is within the wire segment of interest, otherwise it equals zero. Similarly, $\theta$($t_{0}$+$t$) equals one if the molecule remains in the monolayer, and is equal to zero otherwise.  Monolayer molecules were defined as those molecules located a distance from the surface of less than $r_\mathrm{boundary}$, where $r_\mathrm{boundary}$ was taken from results on unstretched wires.  R(t) decays in an exponential fashion, and can thus be written as $\sim$$e^{-t/\tau}$.  The residence time is computed by evaluating:

\begin{equation}\tau = \int_{0}^{t_f} \! R(t) \, dt\label{diffusion}\end{equation} where a delay time, $t_{f}$, of 50 ps was selected to isolate the adsorbate mobility along specific areas of the wire and at particular stages of elongation.  While $\tau$ cannot be directly compared to adsorption rate data, it does enable comparisons for the mobility of the monolayer on different areas of a wire. In Figure 7 (right) the monolayer coverage, $\Theta$, and $\tau$ are shown along different characteristic segments of the wire.  These segments were selected such that the behavior of the adsorbate could be investigated as a function of the local structure along the AuNW surface. 

Figure 7 reveals many interesting features of the adsorbed phase on an elongating AuNW.  Prior to elongation, the monolayer in Figure 7 (left) is highly ordered, with densities as much as twenty times higher than that of the bulk fluid.  Two additional layers of radially ordered molecules are present beyond the monolayer.  As the AuNW is elongated, the monolayer becomes more diffuse and the layers of propane beyond the monolayer become increasingly faint. At 30 {\AA} (immediately prior to rupture of the wire) the adsorbate density is significantly lower than the density prior to elongation.  This drop in monolayer density is caused by the rapid movement of Au atoms in the thinning region, which disrupts the monolayer structure.  

Analyzing the local coverage and mobility of the monolayer in Figure 7 (right) provides further insight into the behavior of the adsorbate.  First, note that $\Theta$ and $\tau$ remain steady in regions where no large structural changes have occurred within the wire. Even slight structural changes, such as the slip planes present at 10 {\AA} elongation, do not significantly affect $\Theta$ or $\tau$.  The region that is most significantly affected is the thinning segment of the wire at 30 \AA of elongation.  Within this region the surface coverage has jumped to 0.72 due to increases in curvature of the AuNW, while the residence time has dropped from 48 ps prior to elongation to 41 ps, a decrease of $\sim$15$\%$.  Thus, the desorption rate is higher in the region surrounding the thinning segment of the wire than in areas surrounding bulk-like regions of the AuNW.

To further probe this phenomena, $\tau$ was calculated for molecules within the thinning region of elongating 1.9 nm AuNWs and compared to $\tau$ for molecules adsorbed along bulk-like regions of the wire.  Ten runs for different adsorbate models were performed.  A 1nm segment in the thinning region was selected for each run, 2 \AA\ prior to rupture of the wire.  Results are presented in Figure 8.  For the LJ/Prop-X models, the monolayer mobility around all regions of the wires gradually increases until X=3.0; the mobility does not change significantly from X=3.0 to X=4.0.  For X=3.0 and X=4.0, the interaction energy is high enough to keep virtually all monolayer molecules attached to bulk regions of the wire for 50 ps.  The explicit models of propane exhibit similar trends.  The mobility of the AA/Prop-UFF model is much higher than that of other models due to its weak interaction with Au.  AA/Prop-FCC has a slightly lower mobility than a LJ/Prop-X model with similar interaction strength.  This finding is consistent with the results for mobility on an unstretched wire, and is attributed to molecular orientation effects. 

\subsection{Impact of Nanowire Thinning Region on Adsorbate Behavior}

Another interesting feature from Figure 8 is the consistently higher mobility of adsorbed molecules in neck versus bulk-like regions of the wires.  The large fluctuations of Au atoms within the thinning region of the wire is the primary factor responsible for the high monolayer mobility. To demonstrate this, in Figure 9 the root-mean-square deviation ($RMSD = \sqrt{\big\langle(r(t)-r_{avg})^2\big\rangle}$) is plotted along the long axis of a 1.9nm AuNW.  The system was allowed to evolve (without stretching of the AuNW) for 5 ns in the presence of different LJ/Prop-X models.  The $RMSD$ for each atom was computed relative to its average position ($r_{avg}$) during the 5 ns trajectory, with the atomic positions saved every 1 ps.   

Figure 9 shows that the fluctuations in Au atomic positions are highest in the thinning region of the wire.  This indicates that the positions of Au atoms within the thinning region are not as tightly bound as Au atoms in bulk-like regions of the wire, as caused by both the bulk motion of the AuNW and local atomic rearrangements.  The large fluctuations in Au atomic positions are likely responsible for the high monolayer mobility surrounding the thinning region.

Outside of the thinning region, the $RMSD$ changes in a linear fashion due to the bulk motion of the AuNW.  Results from Figure 7 and Figure 8 suggest that this bulk motion of the AuNW does not greatly affect the mobility or structure of the monolayer.  That is to say, the monolayer properties remain approximately uniform along bulk-like regions of the AuNW during elongation, in spite of the bulk wire motion.    

Finally, aside from the LJ/Prop-4.0 system, the $RMSD$ curves in Figure 9 are not a strong function of the adsorbate-AuNW interaction strength. The $RMSD$ along the thinning region for LJ/Prop-4.0 is slightly lower than all other systems, indicating that the mobility of these Au atoms is constrained by the strong adsorbate.  However, the bulk motion of the AuNW does not seem to be strongly influenced by high adsorbate-AuNW interaction strengths.  It is possible that this situation may change for heavier molecules, but we did not test such effects here.                         

\subsection{Mechanical and Structural Behavior of Elongating AuNWs}

The rupture of elongating AuNWs is a stochastic process.  Even for AuNWs with the same initial configuration, random fluctuations in atomic trajectories can lead to much different structural pathways during elongation.  Thus, to get a comprehensive picture of the mechanical and structural evolution of elongating AuNWs, ten independent (i.e., each using different initial atomic velocities scaled around 298 K) simulations were run for various combinations of wire size and adsorbate.  Results are compared to those obtained for elongation in vacuum, which are based on 50 independent runs for each wire size.  The mechanical stability of AuNWs was quantified by measuring the ductile elongation of each run. Ductile elongation is defined here as the total elongation that occurs prior to rupture.  

To correlate structural features to the enhanced mechanical stability of AuNWs in adsorbates, the presence of high-energy structural motifs was measured by calculating the diameter in the thinning region of elongating AuNWs.  The thinning region of a AuNW is the location where the thickness of the wire is smallest; as such, it is also the area where rupturing of the wire occurs.  The confinement of Au atoms to the thinning region leads to the frequent appearance of characteristic structural motifs such as monatomic chains (MACs) and helices.  The distinct sizes of these two structures make their appearance straightforward to measure by calculating the diameter of the thinning region.  Helix diameters span between 4.5 and 5.25 \AA\ whereas MAC diameters are equal to the atomic diameter of a single Au atom, $D_\mathrm{Au}$ = 2.88 \AA.  The lengths of MACs and helices may vary.  In their most basic form, MACs represent a single point-contact between two Au atoms.  Longer MACs containing one or more atoms with a coordination number of two can also form.  In either case, such an atomic configuration leads to a conductance through the AuNW equal to the conductance quantum (i.e., G$_o$ =  $\frac{2e^{2}}{h}$ = 77.5 $\mu$S, where $e$ is the charge of an electron and $h$ is Planck's constant) due to the number of conductance channels being reduced to a single channel \cite{Scheer:1998}.  This behavior enables experimentalists to detect the presence of MACs by collecting conductance data through AuNWs during the elongation process (e.g., see Ref. \cite{Yanson:1998}).  We detected the presence of MACs from a large set of AuNW trajectories by calculating the average diameter along a $D_\mathrm{Au}$ = 2.88 \AA\ segment of the wire (in the direction of stretching, [100]). The presence of helices was similarly measured by calculating the average diameter along a 3.5*$D_\mathrm{Au}$ = 10.08 \AA\ segment.

\subsubsection{Ductile Elongation}

Results for ductile elongation as a function of adsorbate-AuNW interaction strength are presented in Figure 10.  The adsorbate-AuNW interaction energy was computed prior to elongation for fixed AuNWs.  Figure 10 shows that the ductile elongation tends to increase as the adsorbate-AuNW interaction energy is increased, irrespective of wire size.  This trend is most noticeable for the 1.1nm AuNW, as the average ductile elongation changes from 21.4 \AA\ in vacuum (i.e., where adsorbate-AuNW energy is zero) to 70.9 \AA\ in the AA/Prop-FCC adsorbate, marking a 231\% increase in mechanical stability.  The enhancements in mechanical stability are smaller for the larger AuNWs.  For instance, the change from 31.7 to 48.8 \AA\ marks a 54\% increase in mechanical stability for the 1.9nm AuNW.  The larger wires also require a higher adsorbate-AuNW interaction energy in order to yield significant improvements in mechanical stability.  This is because larger wires contain a greater fraction of atoms within the core of the wire whose energies are not strongly influenced by the adsorbate.  The surface coverage on the 1.9nm AuNW is also significantly lower than the coverages for the smaller wires (see Figure 5), so Au atoms on the surface of the larger wire interact with fewer adsorbate molecules.  As shown in Figure 7 (right), the surface coverage on a large wire remains relatively uniform on all domains of the wire (excluding the thinning region) during elongation.  While these domains do not exhibit large differences in surface coverage, surface structures appearing on only one side of a wire could lead to radial anisotropy in the ductility enhancement effect.  Our data do not reveal such behavior in these systems; however, we do not rule out the possibility of radial anisotropy-induced effects.  More detailed studies are needed in order to resolve this possibility.    

While the average ductile elongation increases for higher interaction energies, the reproducibility of mechanical stability decreases with larger values of adsorbate-AuNW interaction energy.  Although an adsorbate increases the probability of reaching longer elongations, the stochastic nature of AuNW rupture still results in the occasional breakage at low values of elongation.

\subsubsection{Behavior of Monatomic Chains}

Results from performing analysis of the diameter in the thinning region of AuNWs reveal the appearance of numerous high-energy structures.  MACs form in over 90\% of all simulations, both in vacuum and in adsorbate, irrespective of wire size and adsorbate-AuNW energy.  The fact that MACs form with such high probability is consistent with experimental studies of AuNWs elongated in the [100] direction \cite{Rodrigues:2000}.  Most of the MACs are short in length ($\sim$1-2 atoms) and no correlation between the MAC length and adsorbate-AuNW energy is observed.  MACs of length greater than 3.5*$D_\mathrm{Au}$ = 10.08 \AA\ occur rarely, forming in less than 3\% of all runs.

Experimental data from Ref. \cite{He:2002} indicate that the presence of an adsorbate leads to higher MAC stability.  However, it is unclear whether the adsorbate results in longer MACs or simply prolongs the ``lifetime'' of the structures.  Our results support the latter explanation.  Although they do not result in longer MACs, adsorbates do tend to increase the mechanical stability of MACs.  This can be seen in Figure 11, where histograms of MAC stability in vacuum and in LJ/Prop-3.0 adsorbate are shown.  MAC stability was measured by tracking the amount of elongation that occurred while a MAC was present.  The presence of a strong adsorbate shifts the distribution of MAC stability toward higher elongations.  For example, 80\% of the runs in vacuum result in MAC breakage before we elongate an additional 1.2 \AA, compared to just 10\% of the runs in LJ/Prop-3.0.  The presence of an adsorbate also tends to widen the distribution of MAC stability.  This effect is demonstrated through the large error bars in Figure 12, reflecting the high sensitivity of MAC breakage to thermal effects and adsorbate collisions with the wire.  Figure 12 also shows that the average mechanical stability of MACs tends to increase with higher adsorbate-Au interaction strengths. For a 1.9nm AuNW, the average MAC elongation increases from 0.8 \AA\ in vacuum to 2.4 \AA\ in a strong adsorbate.  These values compare well with experimental results ($\sim$1-3 \AA) for AuNW elongation in the presence of a mercaptopropionic acid monolayer \cite{He:2002} and toluene \cite{Tsutsui_nl:2009}.

\subsubsection{Behavior of Helices}

Further analysis of the thinning region of elongating AuNWs in adsorbate reveals the appearance of a large number of helical structures.  An example of such a helical structure appearing in a 1.9nm AuNW is shown in Figure 13. Helical structures have been experimentally observed in AuNWs by multiple research groups.  For instance, helical core-shell wires with diameters around 0.6 nm have been fabricated and observed \cite{Oshima:2003,Kondo:2000}.  Additionally, rod-like helical structures with diameters just under 2 nm have been observed in elongating AuNWs \cite{Coura:2004,Rodrigues:2000}.  Neither of these structures are thought to be identical to the structure in Figure 13; however, the energetic features are likely similar.

The statistics of the mechanical stability of helical structures in 1.9nm AuNWs are shown in Figure 14.  Helix stability was measured by tracking the amount of AuNW elongation that occurred with at least one helical structure present.  The probability of forming a helical structure is already high in vacuum (78\%).  However, the probability is consistently higher for AuNWs in adsorbate.  Most notably, helices form for LJ/Prop-1.0 and LJ/Prop-3.0 in all ten runs.  In addition to forming helical structures more often, the helices that form in the presence of an adsorbate also possess mechanical stability in excess of those that form in vacuum (see Figure 14).  In fact, the average mechanical stability increases with adsorbate-Au interaction strength, with the changes being more significant after $\epsilon_{Ads.-Au}$/$\epsilon_{Ads.-Ads.}$ exceeds a value of 2.0.  The average helix elongation (1.5-5.8 \AA) is higher than the average MAC elongation (0.8-2.4 \AA), indicating that helical structures possess higher mechanical stability than MACs.  The helices that form in adsorbate are also, on average, longer than helices forming in vacuum.  For example, while the helices that form for a 1.1nm AuNW have an average length of 8.2 \AA, those forming in AA/Prop-FCC have an average length of 11.7 \AA.  For a 1.5nm AuNW, the increase in average helix length goes from 6.6 \AA\ in vacuum to 10.2 \AA\ in LJ/Prop-4.0.  Similarly, the increase goes from 6.2 \AA\ in vacuum to 9.1 \AA\ in LJ/Prop-4.0 for 1.9nm AuNWs.   

\subsubsection{Energetic Considerations}

The mechanism for the enhanced mechanical stability of MACs and helical structures is most easily understood by considering the pertinent energetic factors leading up to AuNW rupture.  As a AuNW is elongated, clusters of Au atoms must rearrange themselves to relieve the strain that is induced through stretching of the wire.  If the strain becomes too high, and atoms are unable to rearrange themselves quickly enough to counter this strain, the wire ruptures.  Atoms on the surface of the wire are further destabilized since they reside in low-coordination environments resulting in higher energies. For this reason, low-coordination structures are especially prone to rupture.  If an adsorbate is present during elongation, it can reduce the energies of surface atoms through strong electronic interactions (e.g., dispersion interactions or covalent bonding), thereby reducing the probability of rupture and possibly prolonging the AuNW ``lifetime.''  This effect is shown in Figure 15, where the average potential energy acting on each Au atom is plotted along the thinning region of an elongating 1.5nm AuNW.  Two scenarios are plotted: the top curve shows the potential energy per particle in vacuum, while the bottom curve includes the contributions of both the Au-Au interactions and the adsorbate-Au interactions.  The top curve was obtained by removing the adsorbate molecules from the simulation box and allowing the wire to evolve in vacuum without stretching.  Both curves represent the averages of configurations taken every 50 fs for 50 ps.  The adsorbate reduces the potential energy acting on Au atoms considerably, especially in areas of low Au coordination.       

The effect of an adsorbate on the appearance of high-energy structural motifs is further demonstrated by calculating the Au-Au interaction energy immediately prior to the rupture of 1.9nm AuNWs.  The Au-Au potential energy is plotted as a function of adsorbate-AuNW interaction energy in Figure 16.  The Au-Au energy increases exponentially with adsorbate-AuNW interaction strength.  In other words, AuNWs are able to adopt and maintain unfavorable atomic configurations better in the presence of an adsorbate than in the absence of one.  This effect occurs not only because the interaction between the adsorbate and AuNW is stronger, but also by virtue of the fact that higher surface coverages result from higher interaction strengths, providing Au atoms with more adsorbate molecules to interact with.  This result provides further evidence that the enhanced mechanical properties observed in Figure 10 are the result of the formation of high-energy structural motifs.

\subsubsection{The Effect of a Weakly Interacting Molecular Species on MAC Stability}

The total energy associated with each Au atom is not the only factor that influences AuNW breakage.  The forces acting on each atom can also play an important role.  For example, a molecular species that does not interact strongly with a AuNW may destabilize high-energy structural motifs through the bombardment of molecules onto the AuNW surface.  Previous work in our group found that such a solvent species does not affect the overall ductile elongation of AuNWs \cite{Pu:2007}.  Although the overall ductile elongation is not affected by a solvent, the mechanical stability of high-energy structures such as MACs may still be influenced.  To test this possibility, the MAC mechanical stability was measured in vacuum and in AA/Prop-UFF.  AA/Prop-UFF was selected because its $\epsilon_{Ads.-Au}$ value is the lowest among the adsorbate models tested. The results for three different wire sizes are shown in Figure 17.  The distributions of MAC stability in vacuum and in AA/Prop-UFF are similar at MAC elongations of less than 2 \AA.  However, the situation changes at higher elongations.  At elongation lengths greater than 2 \AA, MAC breakages occur occasionally in vacuum but never occur in AA/Prop-UFF.  This tightening in the MAC stability distribution is reflected in the standard deviations for MAC mechanical stability relative to the in-vacuum runs.  The standard deviations of MAC elongation for 1.1, 1.5, and 1.9nm AuNWs are 2.31, 0.76, and 0.57 \AA\ in vacuum and 0.43, 0.40, and 0.27 \AA\ in AA/Prop-UFF, respectively.  While MACs are at times able to sustain themselves in vacuum up to high elongations, the bombardment of molecules onto the AuNW surface prevents this from occurring in AA/Prop-UFF.  

This is a somewhat surprising result since the mass of propane (44.1 g/mol) is low relative to a Au atom (197.0 g/mol), and illustrates the high instability of MACs.  We expect these findings to be even more apparent for heavier molecules that interact weakly with Au and have a relatively high monolayer mobility.  The high monolayer mobility occurring in the thinning region of AuNWs (see Figure 8) may additionally help promote MAC breakage by increasing the frequency of solvent collisions with the AuNW surface.    

\section{Conclusions}

We have performed molecular dynamics simulations to test the effect of molecular adsorption on the mechanical and morphological evolution of elongating AuNWs.  A simple Lennard-Jones sphere model and a fully atomistic model for the adsorbate were tested on three different wire sizes (1.1, 1.5, and 1.9 nm in diameter).  With each adsorbate model the adsorbate-Au interaction strength was systematically altered to get a comprehensive picture of the effect of different adsorbates on elongating AuNWs.  As the interaction strength is increased, the monolayer surface coverage increases while its mobility decreases.  The process of mechanically elongating a AuNW does not impact the monolayer surrounding bulk-like regions of the wire.  However, along the thinning region of a AuNW, the adsorbed phase is found to be less structured and more mobile than phases located on bulk-like regions of the wire.  These differences are attributed to fluctuations in the atomic positions of Au atoms that occur in the thinning region. 

The ductile elongation of AuNWs in the presence of an adsorbate is enhanced relative to elongation in vacuum, by 231\% for 1.1nm AuNWs and 54\% for 1.9nm AuNWs.  This result is explained by the lowering of the AuNW surface energy due to the presence of an adsorbate, which prolongs the ``lifetime'' of low-coordination structures that are especially prone to rupture in vacuum environments.  The Au-Au energy immediately prior to AuNW rupture is found to increase exponentially with adsorbate-AuNW energy, providing further evidence that high-energy structures are stabilized through molecular adsorption onto a AuNW surface. 

The appearance and stability of structures in the thinning region of elongating AuNWs were also computed. While the presence of an adsorbate does not result in longer monatomic chains, it does improve their mechanical stability.  The MAC stability data we obtain are in good quantitative agreement with experimental data from Refs. \cite{He:2002,Tsutsui_nl:2009}.  On the other hand, molecular species that interact weakly with AuNWs worsen the mechanical stability of MACs.  Helices that form in the presence of adsorbate are found to ($i$) occur with higher frequency, ($ii$) possess higher mechanical stability, and ($iii$) display greater lengths than those that appear in vacuum environments.  

Finally, a simple Lennard-Jones model was found to adequately describe the impact of an adsorbate on the properties of the monolayer and the mechanical stability of AuNWs, with only minor differences noted in monolayer mobility when compared to a fully atomistic adsorbate model.  This is an important result since varying the interaction strength for a LJ model is straightforward and also because the LJ model is a computationally cheaper alternative than an explicit model.  These results should provide guidance for future simulation studies of AuNW elongation in solvent and/or an adsorbing species. 

The trends we observe in this paper suggest that simple adsorbates may be used as a design tool to influence the properties and structures of elongating AuNWs.  Furthermore, our results hint at the importance of atomic-level structural effects that result from adsorbates (e.g., thiol and amine-terminated organic molecules) in molecular break junction experiments.

\section{Acknowledgements}

W. F. acknowledges support from the U.S. Department of Education for a Graduate Assistance in Areas of National Need (GAANN) Fellowship under grant number P200A090323, as well as the U.S. Department of Energy under grant number DEFG0203ER46096.  This research used resources of the National Energy Research Scientific Computing Center, which is supported by the Office of Science of the U.S. Department of Energy, under grant number DOE KC0204010-ERKCZ01.  This work was also supported in part by the National Science Foundation through TeraGrid resources provided by Texas Advanced Computing Center and Oak Ridge National Laboratory, under grant number TG-DMR090099.

\pagebreak

%
%
%
\bibliography{refs_library}

%
%
%
%
%
%

%
%
\begin{table} []
\caption{Force Field Parameters.}
\centering
\begin{tabular}{ c  c  c  c  c}
\multicolumn{5}{c}{Lennard-Jones Parameters and Partial Charges} \\
    \hline 
System \T \B & Atom & $\epsilon_{i}$ & $\sigma_{i}$ & q$_{i}$ \\
 \T \B & & (kcal/mol) & (\AA) & \\
    \hline 
LJ/Prop-X \T \B & - & 0.553 & 4.660 & - \\
AA/Prop-Y \cite{Jorgensen:1996} \T \B & C & 0.066 & 3.500 & -0.24 \\
AA/Prop-Y \cite{Jorgensen:1996} \T \B & H & 0.030 & 2.500 & 0.06 \\
     \hline \\
\multicolumn{5}{c}{Adsorbate-Au Lennard-Jones Parameters for AA/Prop-Y} \\
	\hline
System \T \B & $\epsilon_{Au-C}$ & $\sigma_{Au-C}$ & $\epsilon_{Au-H}$ & $\sigma_{Au-H}$ \\
 \T \B & (kcal/mol) & (\AA) & (kcal/mol) & (\AA) \\
	\hline
AA/Prop-UFF \cite{Jorgensen:1996,Rappe:1992} \T \B & 0.05 & 3.20 & 0.03 & 2.71 \\
AA/Prop-MP2 \cite{Piana:2006} \T \B & 0.10 & 3.83 & 0.04 & 3.53  \\
AA/Prop-FCC \cite{Jorgensen:1996,Heinz:2008} \T \B & 0.59 & 3.03 & 0.40 & 2.56 \\
	\hline \\
	
\multicolumn{5}{c}{AA/Prop-Y Bond Stretch Parameters \cite{Jorgensen:1996}} \\
	\hline
Type \T \B & $r_{eq}$ & $K_r$ & &  \\
 \T \B & (\AA) & ((kcal/mol)/\AA$^2$) & &  \\
	\hline
C-C \T \B & 1.529 & 268.0  & &  \\
C-H \T \B & 1.090 & 340.0  & &  \\
	\hline \\ 
	
\multicolumn{5}{c}{AA/Prop-Y Bond Angle Bending Parameters \cite{Jorgensen:1996}} \\
	\hline
Type \T \B & $\theta_{eq}$ & $K_{\theta}$ & & \\
 \T \B & (deg.) & ((kcal/mol)/rad$^2$) & & \\
	\hline
C-C-C \T \B & 112.7 & 58.35  & & \\
C-C-H \T \B & 110.7 & 37.50  & & \\
H-C-H \T \B & 107.8 & 33.00  & & \\
	\hline \\ 
	
\multicolumn{5}{c}{AA/Prop-Y Torsional Parameters, in kcal/mol \cite{Jorgensen:1996}} \\
	\hline
Type \T \B & $V_{1}$ & $V_{2}$ & $V_{3}$ & \\
	\hline
C-C-C-H \T \B & 0.000 & 0.000  & 0.366 & \\
H-C-C-H \T \B & 0.000 & 0.000  & 0.318 & \\
	\hline \\

\end{tabular}
\label{table:Parameters}
\end{table}  

%
%
\begin{table} []
\caption{Adsorbate-adsorbent well depth energies.}
\centering
\begin{tabular}{ | c | c | c | c |}
	\hline
System \T \B & $\epsilon_{Ads.-Ads.}$ & $\epsilon_{Ads.-Au}$ & $\frac{\epsilon_{Ads.-Au}}{\epsilon_{Ads.-Ads.}}$ \\
 \T \B & (kcal/mol) & (kcal/mol) & \\
	\hline
LJ/Prop-0.5 \T \B & 0.553 & 0.277 & 0.50 \\
LJ/Prop-1.0 \T \B & 0.553 & 0.553 & 1.00 \\
LJ/Prop-2.0 \T \B & 0.553 & 1.106 & 2.00 \\
LJ/Prop-3.0 \T \B & 0.553 & 1.659 & 3.00 \\
LJ/Prop-4.0 \T \B & 0.553 & 2.212 & 4.00 \\
          \hline
AA/Prop-UFF \T \B & 1.686 & 0.286 & 0.17 \\
AA/Prop-MP2 \T \B & 1.686 & 0.447 & 0.27  \\
AA/Prop-FCC \T \B & 1.686 & 2.898 & 1.72  \\
	\hline
\end{tabular}
\label{table:energies}
\end{table}         

%
%
\begin{table} [] 
\caption{Equation 2 fitting parameters for LJ/Prop-X systems.}
\centering
\begin{tabular}{ | c | c | c | }
	\hline
Diameter (nm) \T \B & $\alpha$ & $\beta$ \\
	\hline
1.1 \T \B & 0.571 & 0.299  \\
1.5 \T \B & 0.569 & 0.273  \\
1.9 \T \B & 0.514 & 0.210  \\
	\hline
\end{tabular}
\label{table:coverage}
\end{table}            

\pagebreak
%
%
%
%
\section{Figures}

%
%
\begin{figure}[H]
	\centering
	\includegraphics[width=2in]{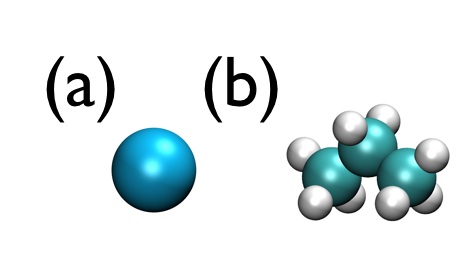}
	\caption{Adsorbate models: (a) Lennard-Jones propane (LJ/Prop-X), (b) all-atom propane (AA/Prop-Y).  Images are drawn to scale.}
	\label{fig:models}
\end{figure}

%
%
\begin{figure}[H]
	\centering
	\includegraphics[width=5.0in]{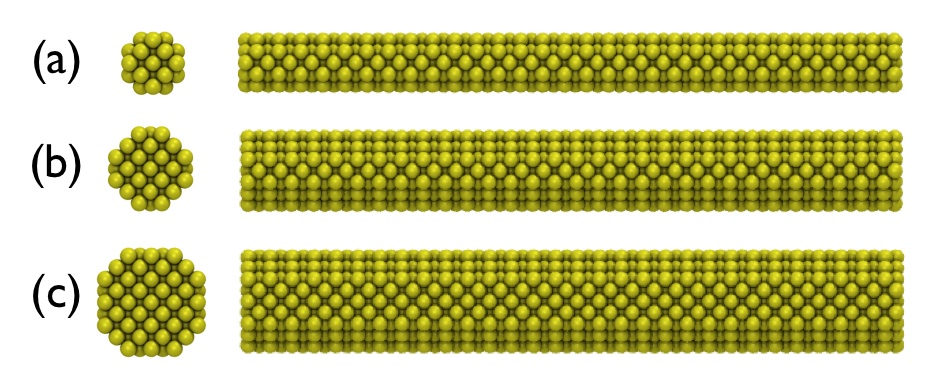}
	\caption{Top (left) and side (right) views of gold nanowires with diameters of (a) 1.1 nm, (b) 1.5 nm, and (c) 1.9 nm.  The wires contain 630, 1110, and 2070 atoms, respectively. }
	\label{fig:NWs}
\end{figure}

%
%
\begin{figure}[H]
	\centering
	\includegraphics[width=5.0in]{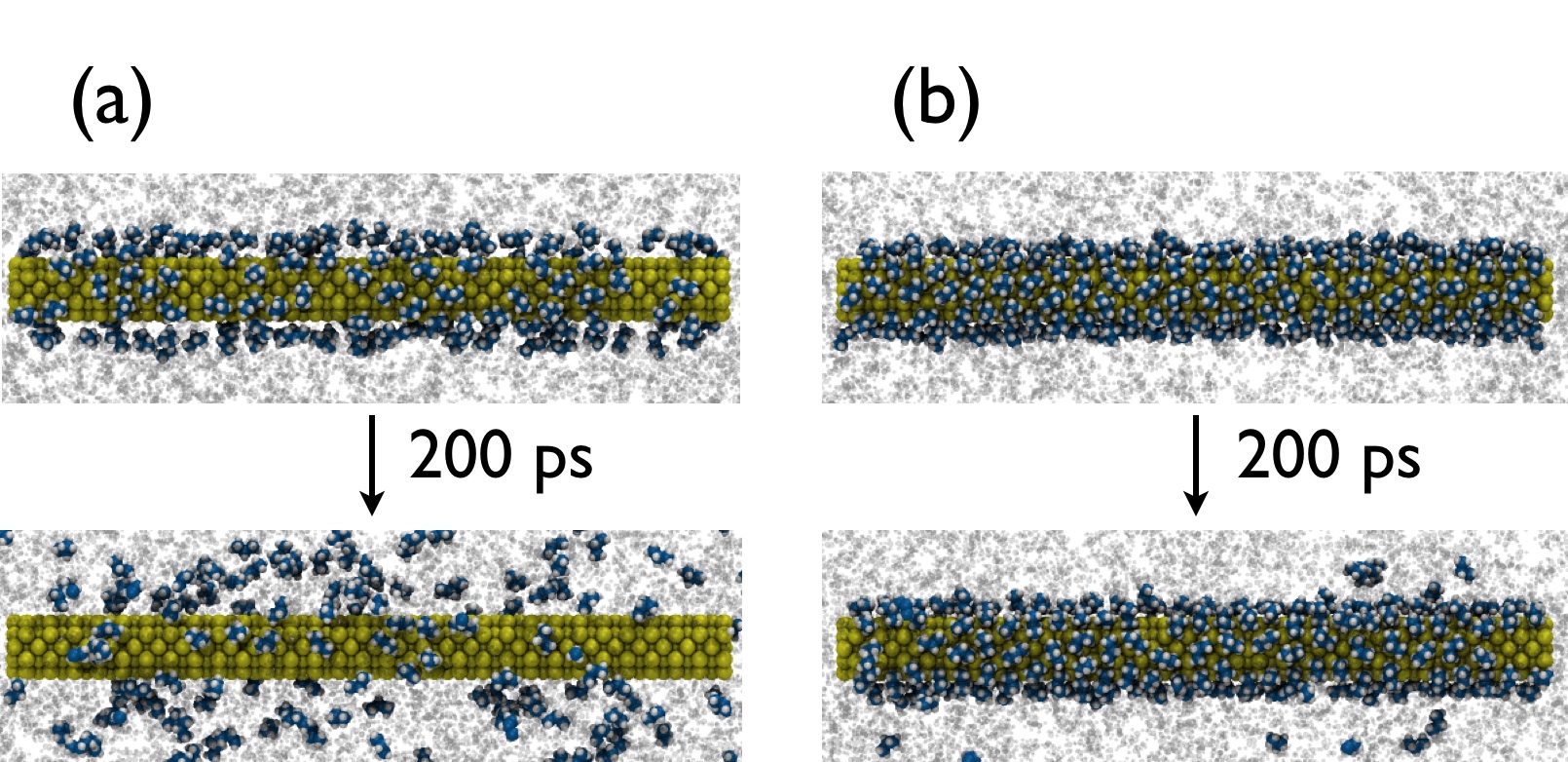}
	\caption{Snapshots depicting the structure and mobility of monolayer molecules around a 1.1nm AuNW for the (a) AA/Prop-MP2 and (b) AA/Prop-FCC models.  The top snapshots show the monolayer molecules (whose atoms are rendered as van der Waals spheres) at a given instant in time, while the bottom snapshots show these same molecules 200 ps later.  Molecules outside of the initial monolayer are rendered as grey ``ghost'' molecules.}
	\label{fig:monolayer}
\end{figure}

%
%
\begin{figure}[H]
	\centering
	\includegraphics[width=5.0in]{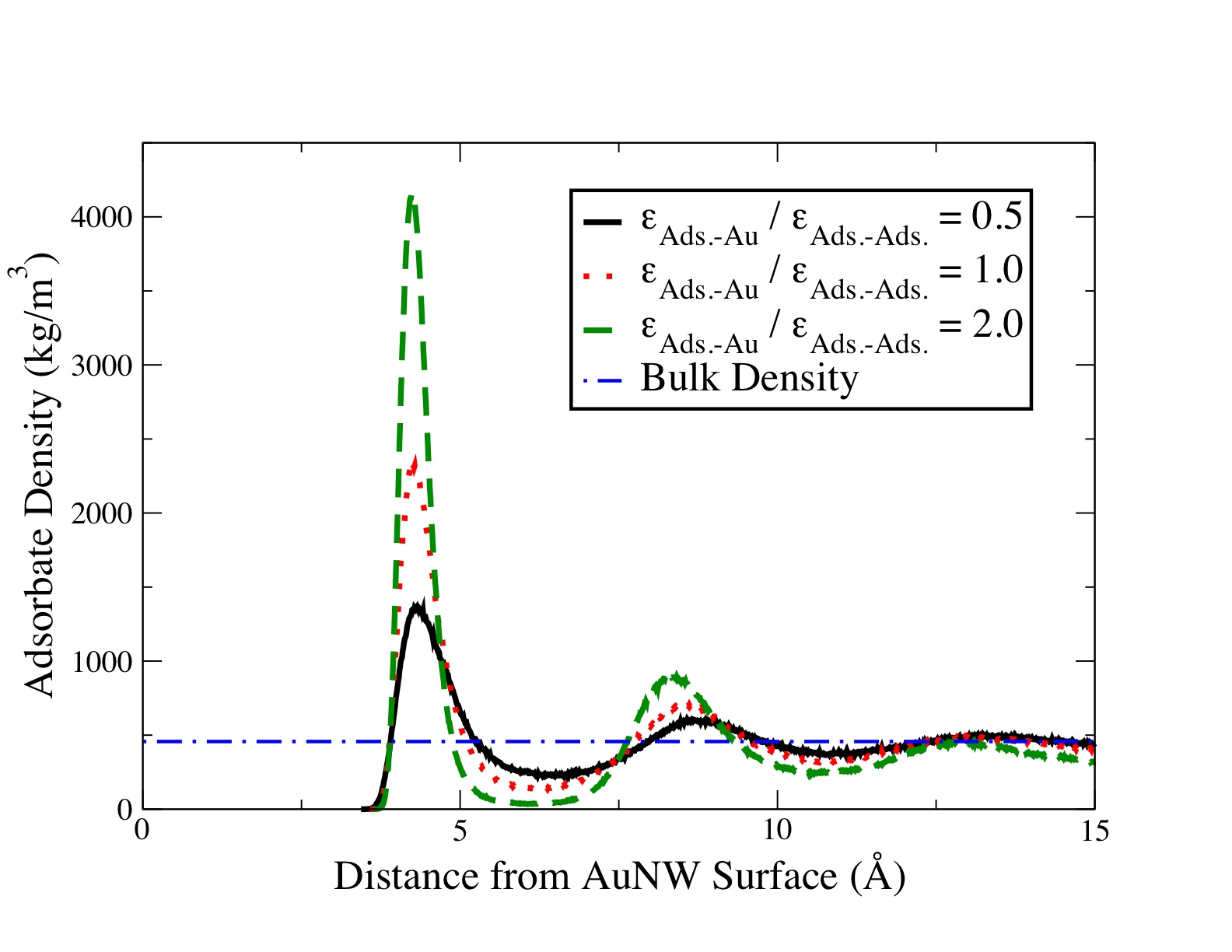}
	\caption{Adsorbate densities around a 1.1nm AuNW for LJ/Prop-X model using three different interaction strengths.}
	\label{fig:rho}
\end{figure}

%
%
\begin{figure}[H]
	\centering
	\includegraphics[width=5.0in]{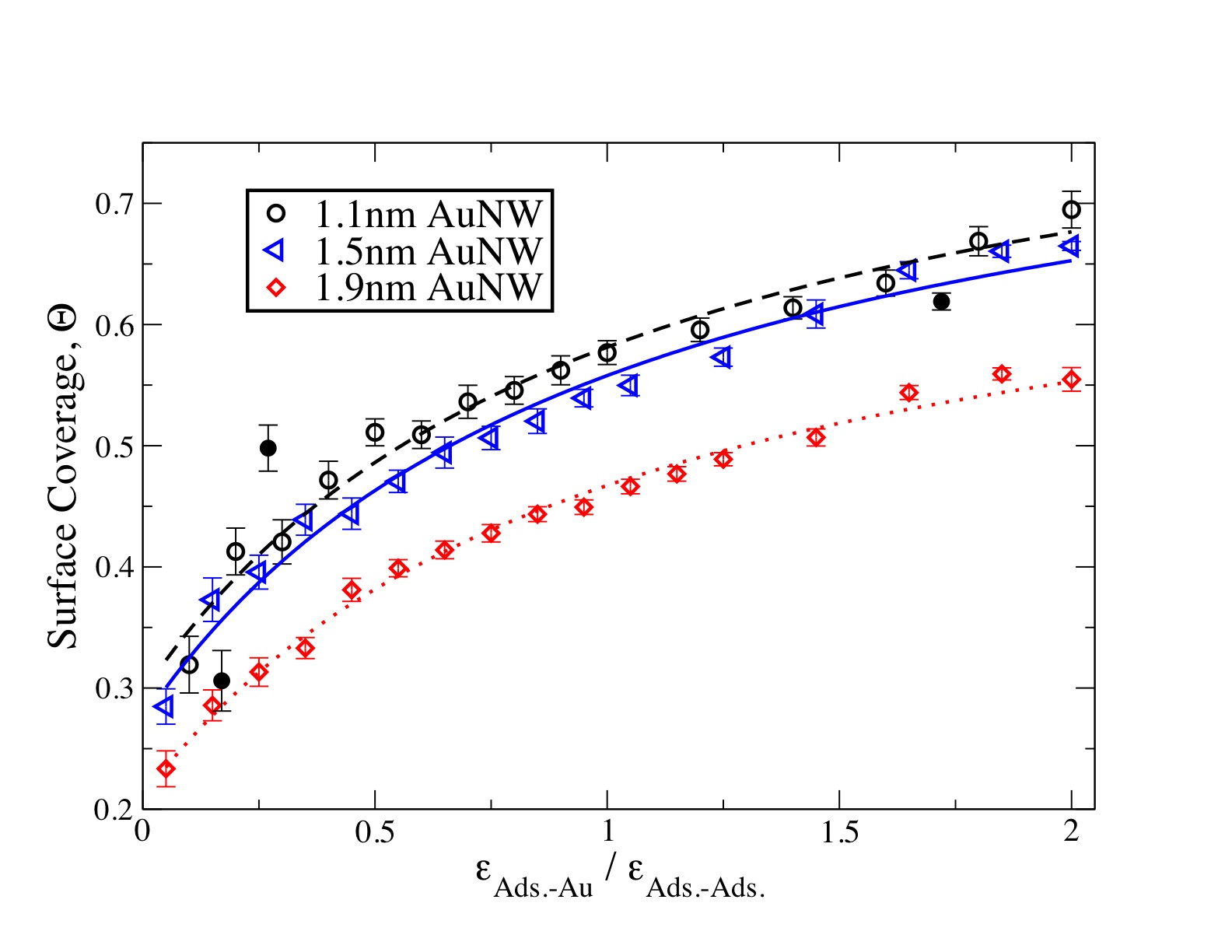}
	\caption{LJ/Prop-X (unfilled symbols) monolayer coverage on 1.1nm, 1.5nm, and 1.9nm AuNWs.  Langmuir isotherm-type curves are fit to each set of LJ/Prop-X data.  AA/Prop-Y data (filled symbols) are also plotted for the 1.1nm AuNW.}
	\label{fig:coverage}
\end{figure}

%
%
\begin{figure}[H]
	\centering
	\includegraphics[width=5.0in]{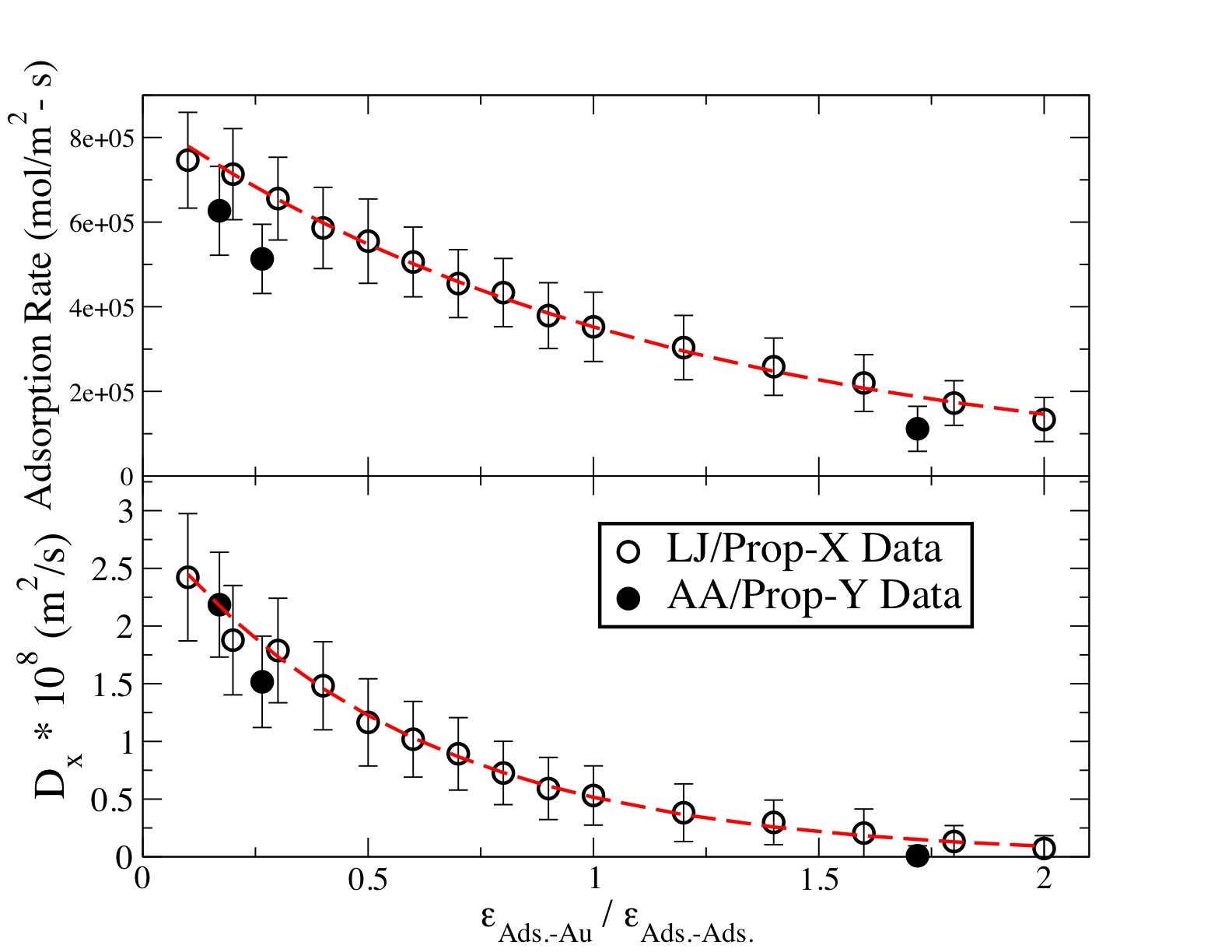}
	\caption{Adsorbate mobility on a 1.1nm AuNW as a function of adsorbate-Au interaction strength.  Exponential fits are applied to the LJ/Prop-X data. (Top) Adsorption rate and (Bottom) diffusion along AuNW surface (in [100] direction) are plotted. }
	\label{fig:mobility}
\end{figure}

%
%
\begin{figure}[H]
	\centering
	\includegraphics[width=4.0in]{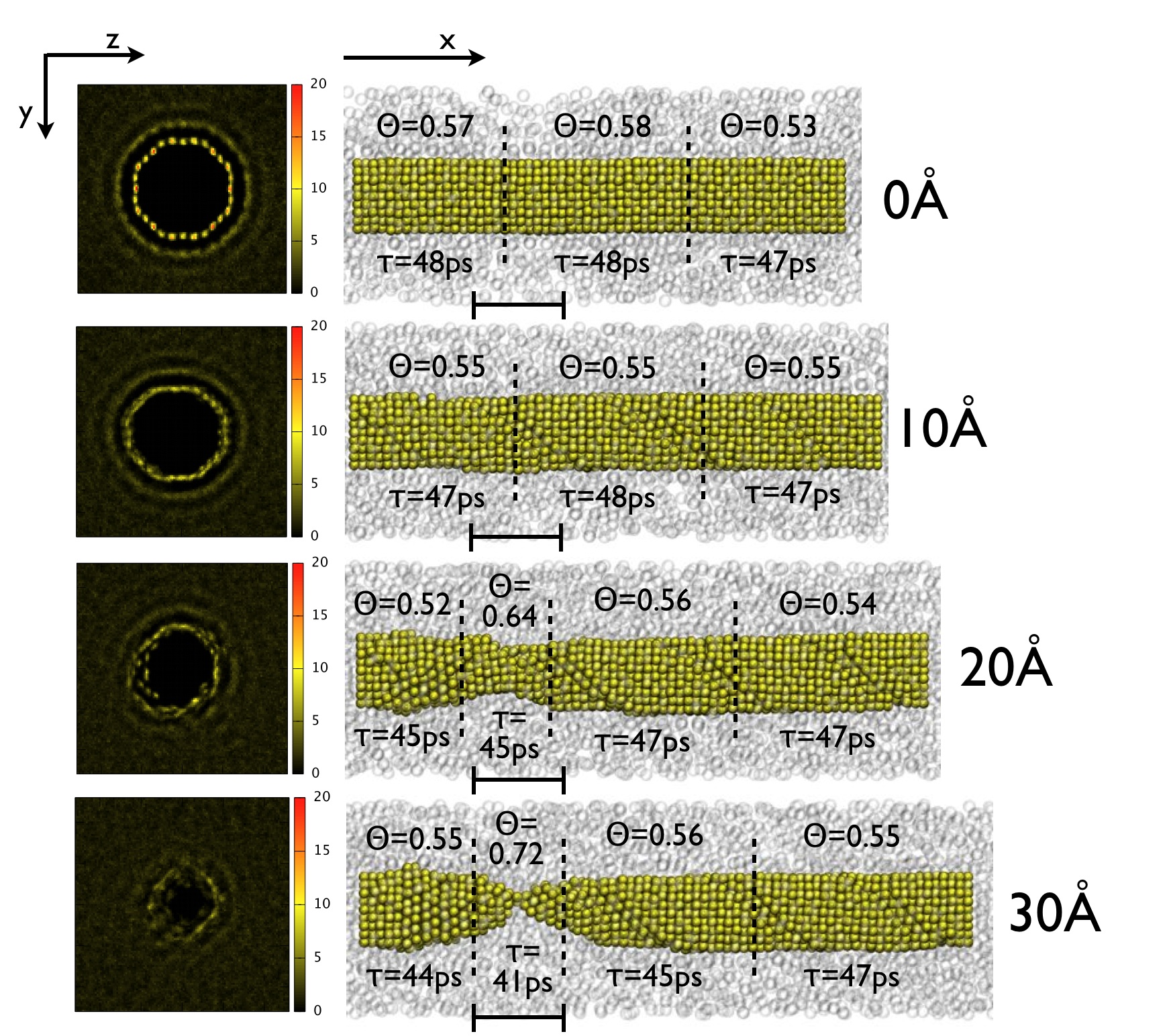}
	\caption{Structure and mobility of LJ/Prop-2.0 adsorbate at different stages of elongation of a 1.9nm AuNW.  (left) Adsorbate density (normalized with respect to the bulk density) along a segment of the wire, from x=3.0 nm to x=5.3 nm.  This segment corresponds to the thinning region enclosed by the solid lines at the 30 \AA\ stage of elongation (right). (right) The surface coverage, $\Theta$, and desorption residence time, $\tau$, along different segments of the elongating wire. }
	\label{fig:dynamic_propane}
\end{figure}

%
%
\begin{figure}[H]
	\centering
	\includegraphics[width=5.0in]{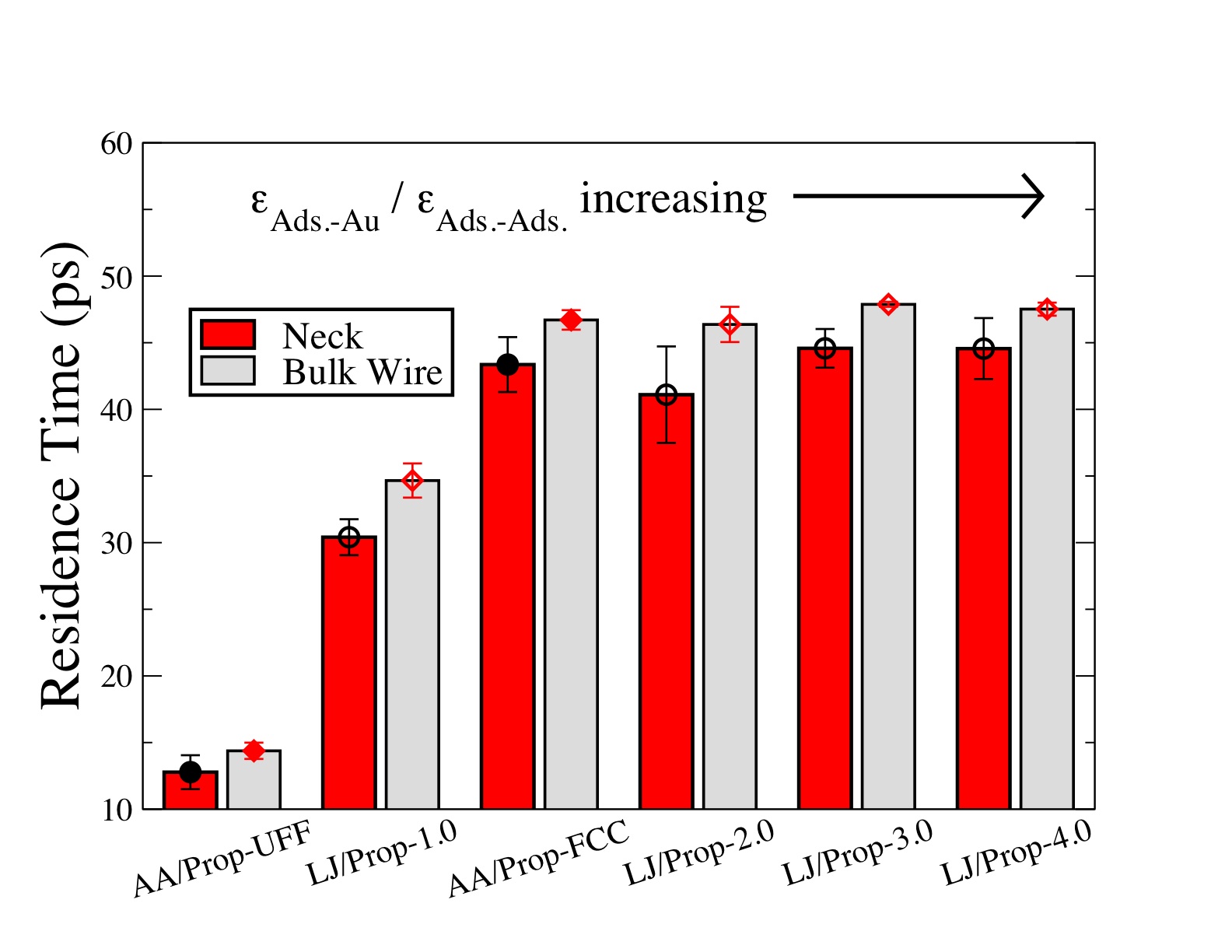}
	\caption{Comparison of various adsorbate desorption residence times at the neck (i.e., the thinning region) and in bulk-like regions of elongating 1.9nm AuNWs. }
	\label{fig:tao}
\end{figure}

%
%
\begin{figure}[H]
	\centering
	\includegraphics[width=5.0in]{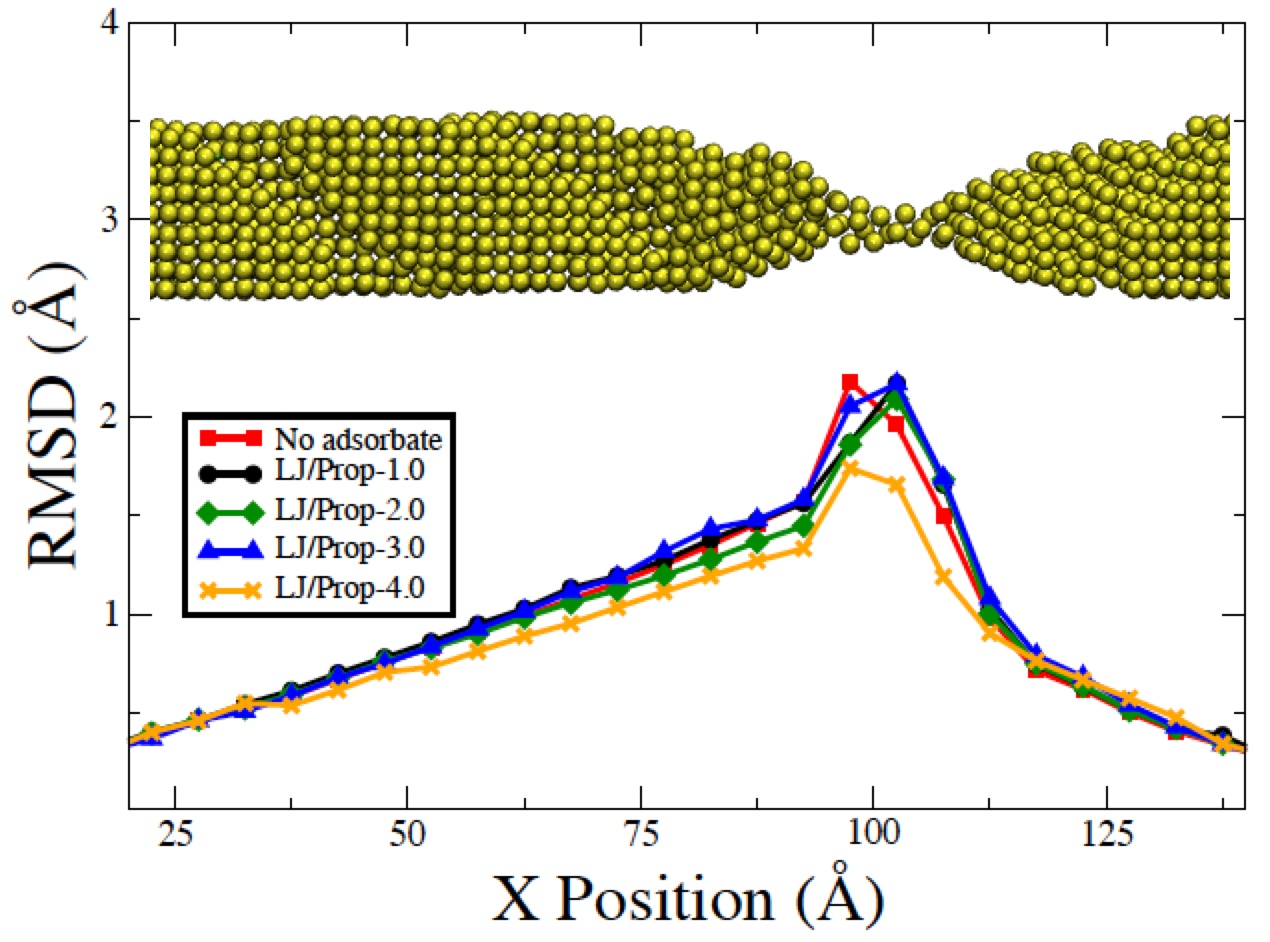}
	\caption{Root-mean-square deviation (RMSD) of Au atomic positions along a 1.9nm AuNW.  Curves for the RMSD in vacuum and in different LJ/Prop-X models are shown.  Note, wire is shown to scale and matches the x-axis. }
	\label{fig:oscillations}
\end{figure}

%
%
\begin{figure}[H]
	\centering
	\includegraphics[width=5.0in]{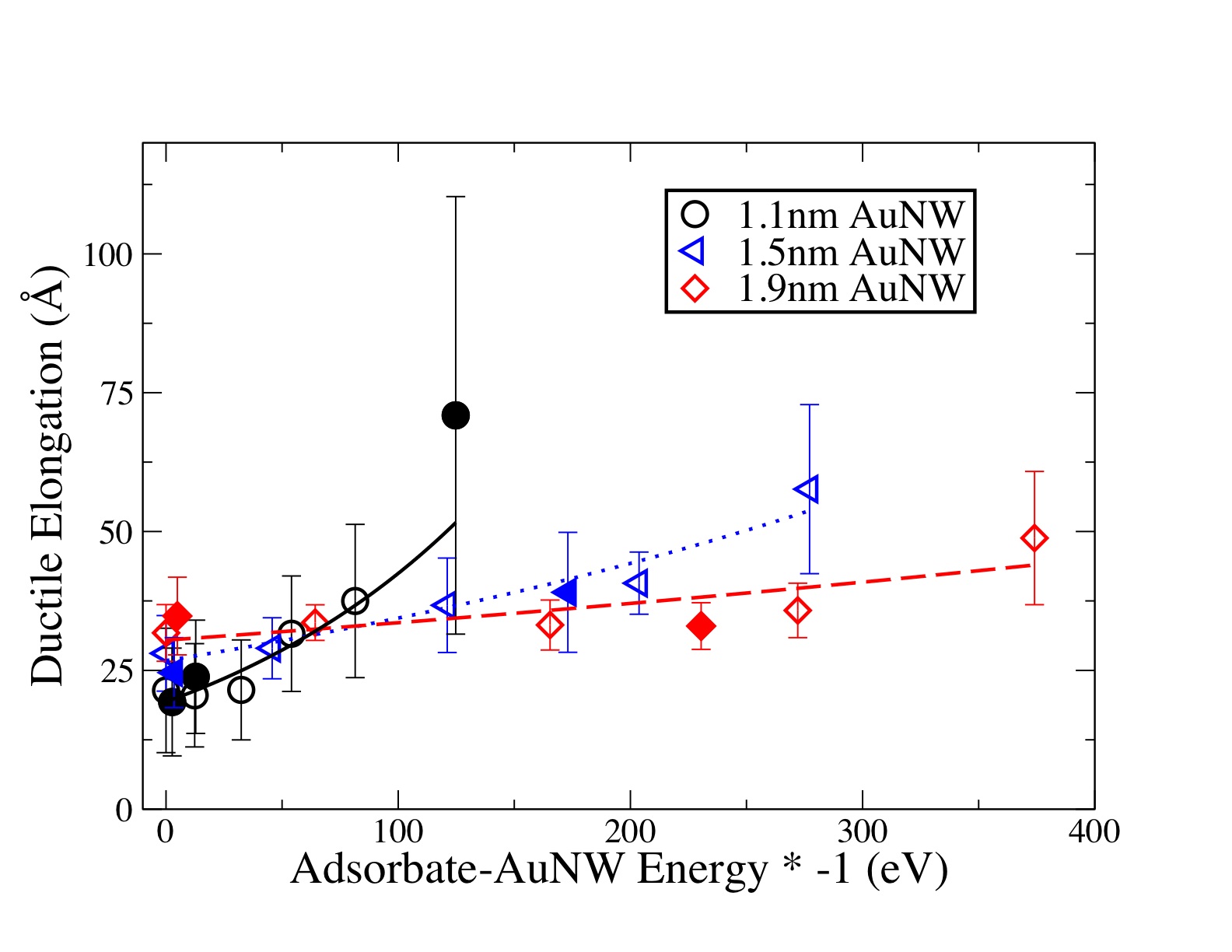}
	\caption{Ductile elongation of three wire sizes as a function of adsorbate-Au interaction energy.  Unfilled symbols correspond to LJ/Prop-X data while the filled symbols represent AA/Prop-Y data.  Exponential fits are applied to LJ/Prop-X data for each wire size. }
	\label{fig:elon}
\end{figure}

%
%
\begin{figure}[H]
	\centering
	\includegraphics[width=5.0in]{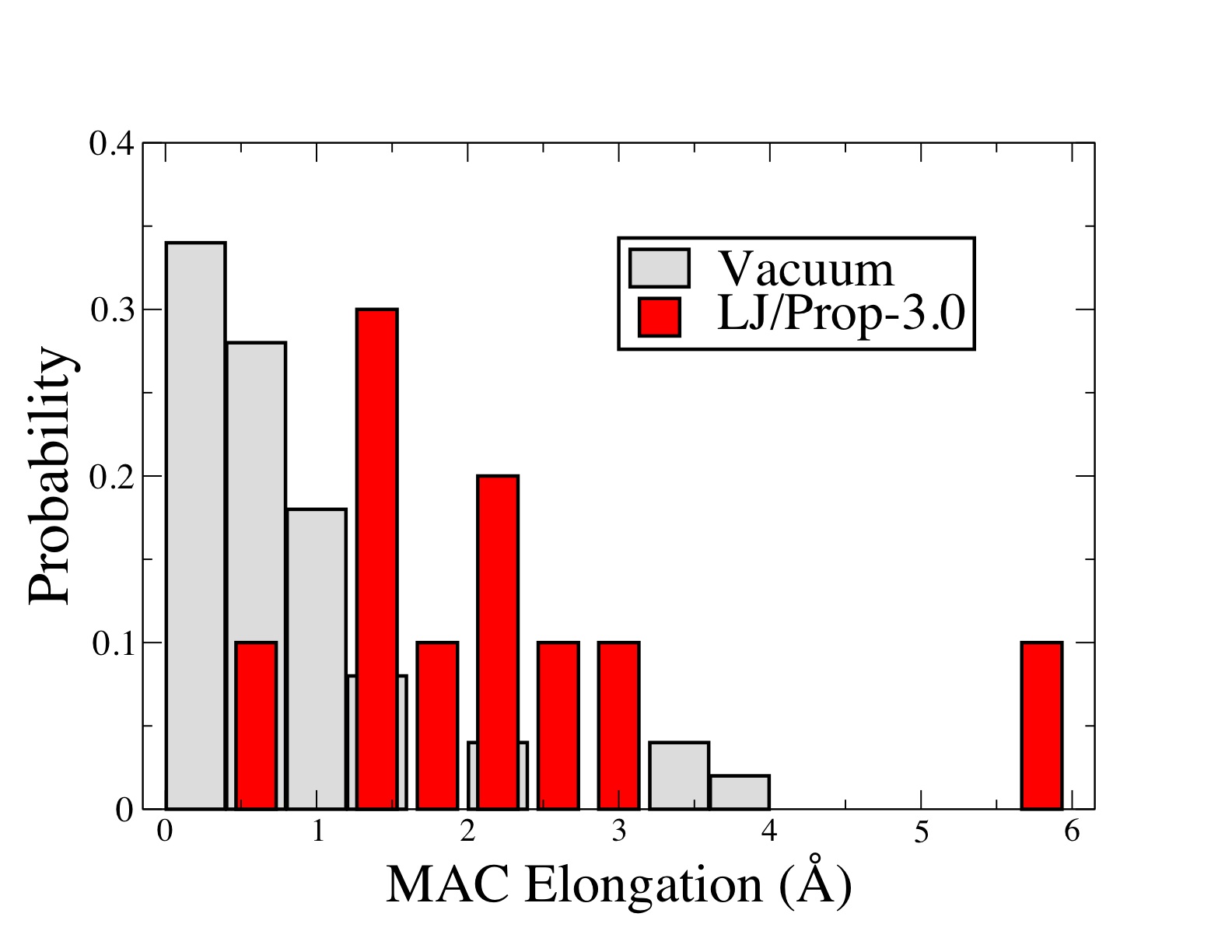}
	\caption{Histograms of monatomic chain stability in vacuum and in LJ/Prop-3.0 adsorbate for 1.9nm AuNWs. }
	\label{fig:stab_histo}
\end{figure}

%
%
\begin{figure}[H]
	\centering
	\includegraphics[width=5.0in]{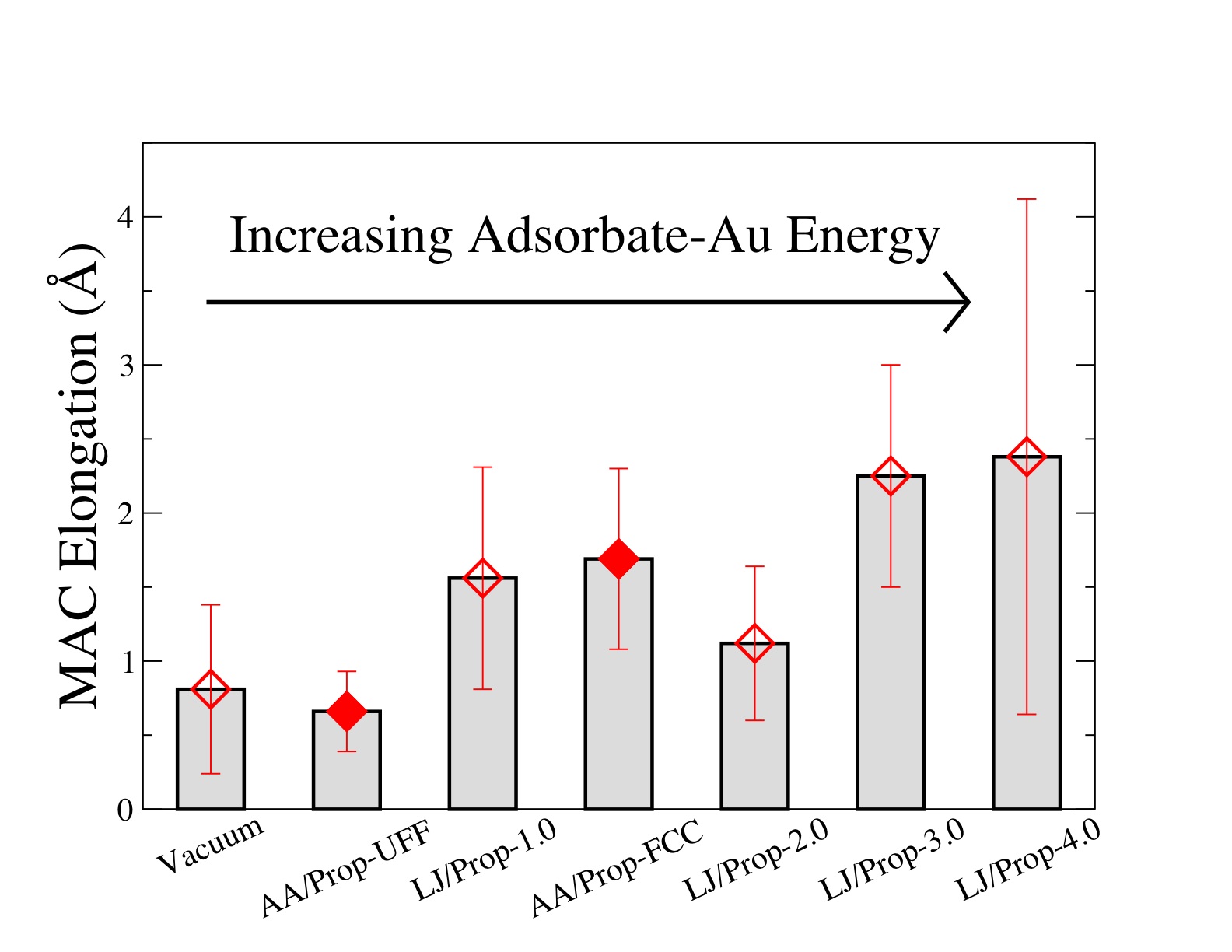}
	\caption{Monatomic chain elongation length in vacuum and in various adsorbates for a 1.9nm AuNW. }
	\label{fig:mac_elon}
\end{figure}  

%
%
\begin{figure}[H]
	\centering
	\includegraphics[width=4.0in]{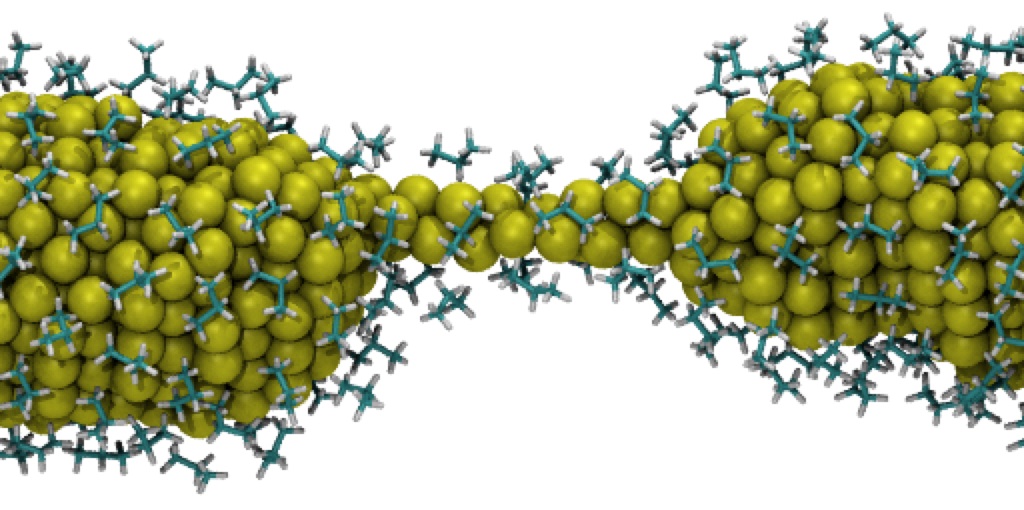}
	\caption{Helical formation in 1.9nm AuNW elongating in AA/Prop-FCC adsorbate. Molecules outside of the monolayer are removed for clarity. }
	\label{fig:helix}
\end{figure}

%
%
\begin{figure}[H]
	\centering
	\includegraphics[width=5.0in]{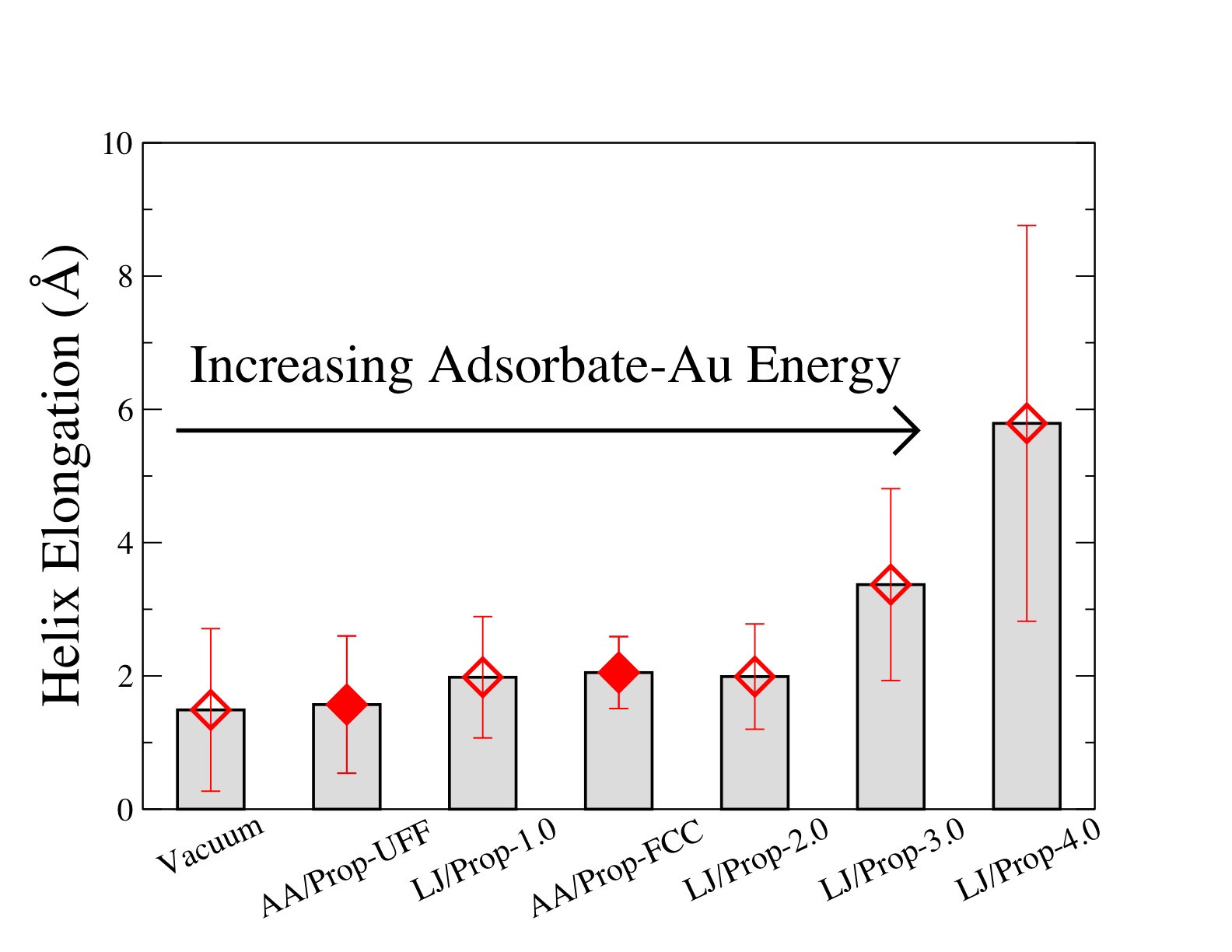}
	\caption{Average helix elongation length in vacuum and in various adsorbates for a 1.9nm AuNW.}
	\label{fig:helix_stab}
\end{figure}   

%
%
\begin{figure}[H]
	\centering
	\includegraphics[width=4.0in]{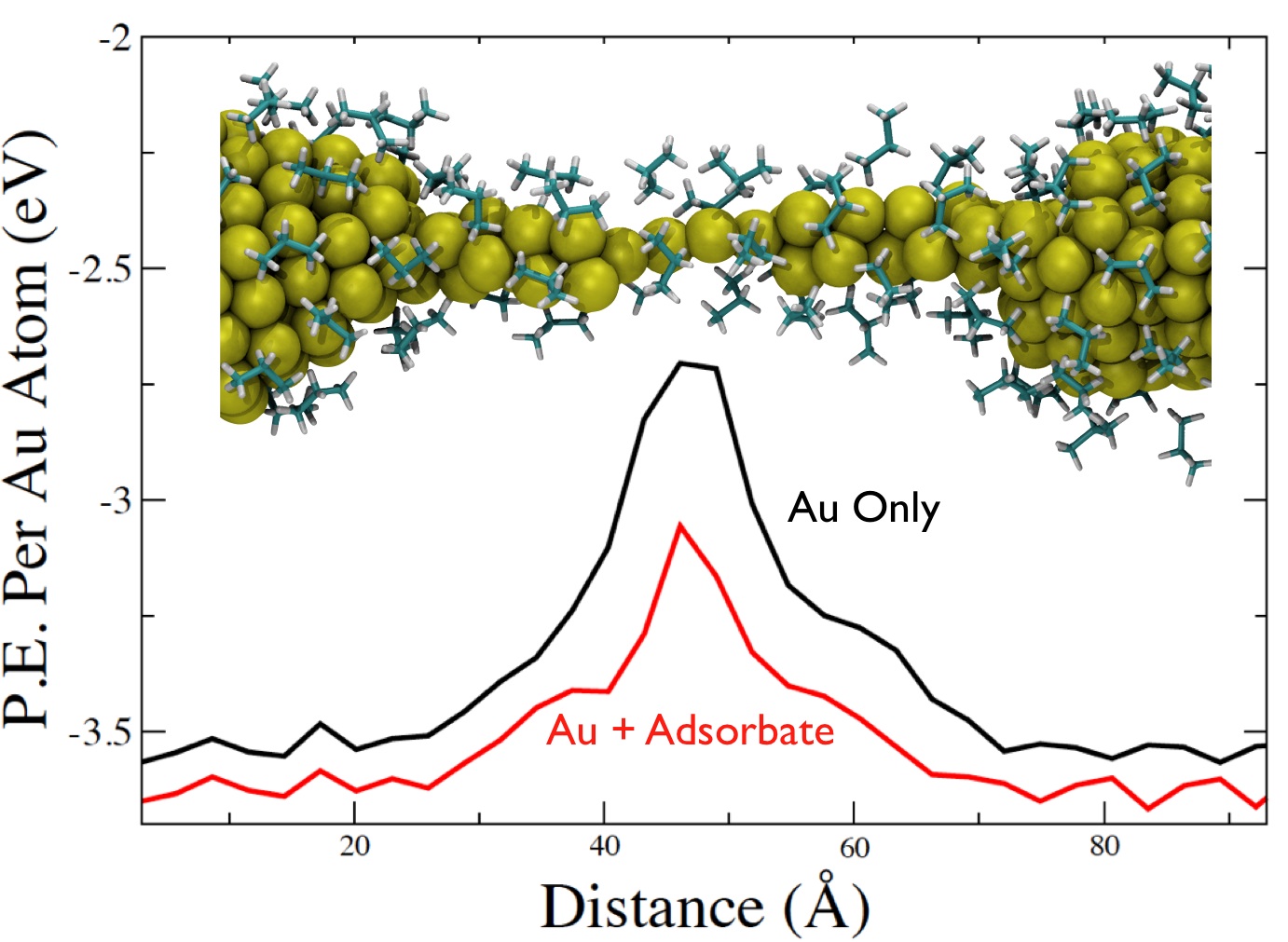}
	\caption{(Inset Image) Thinning region of a 1.5nm AuNW in AA/Prop-FCC.  Molecules outside of the monolayer are removed for clarity.  (Top Curve) The average potential energy acting on each Au atom in vacuum.  (Bottom Curve) The average potential energy acting on each Au atom, including the contribution of both the Au-Au and adsorbate-Au interactions. }
	\label{fig:pe_profile}
\end{figure}

%
%
\begin{figure}[H]
	\centering
	\includegraphics[width=5.0in]{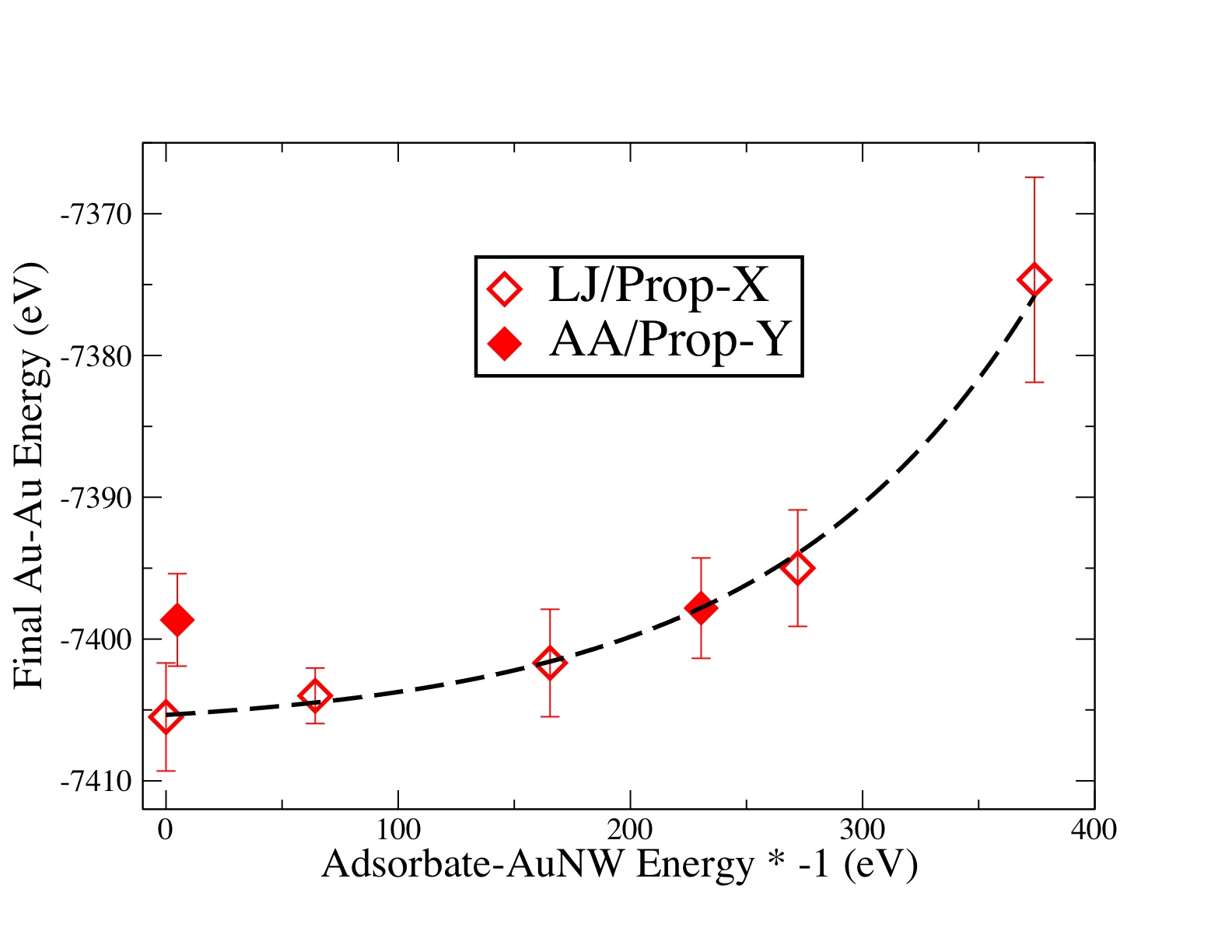}
	\caption{Au-Au energy immediately before rupture as a function of adsorbate-AuNW interaction strength for 1.9nm AuNWs.  An exponential fit is applied to the LJ/Prop-X data.}
	\label{fig:final_eng}
\end{figure}

%
%
\begin{figure}[H]
	\centering
	\includegraphics[width=5.0in]{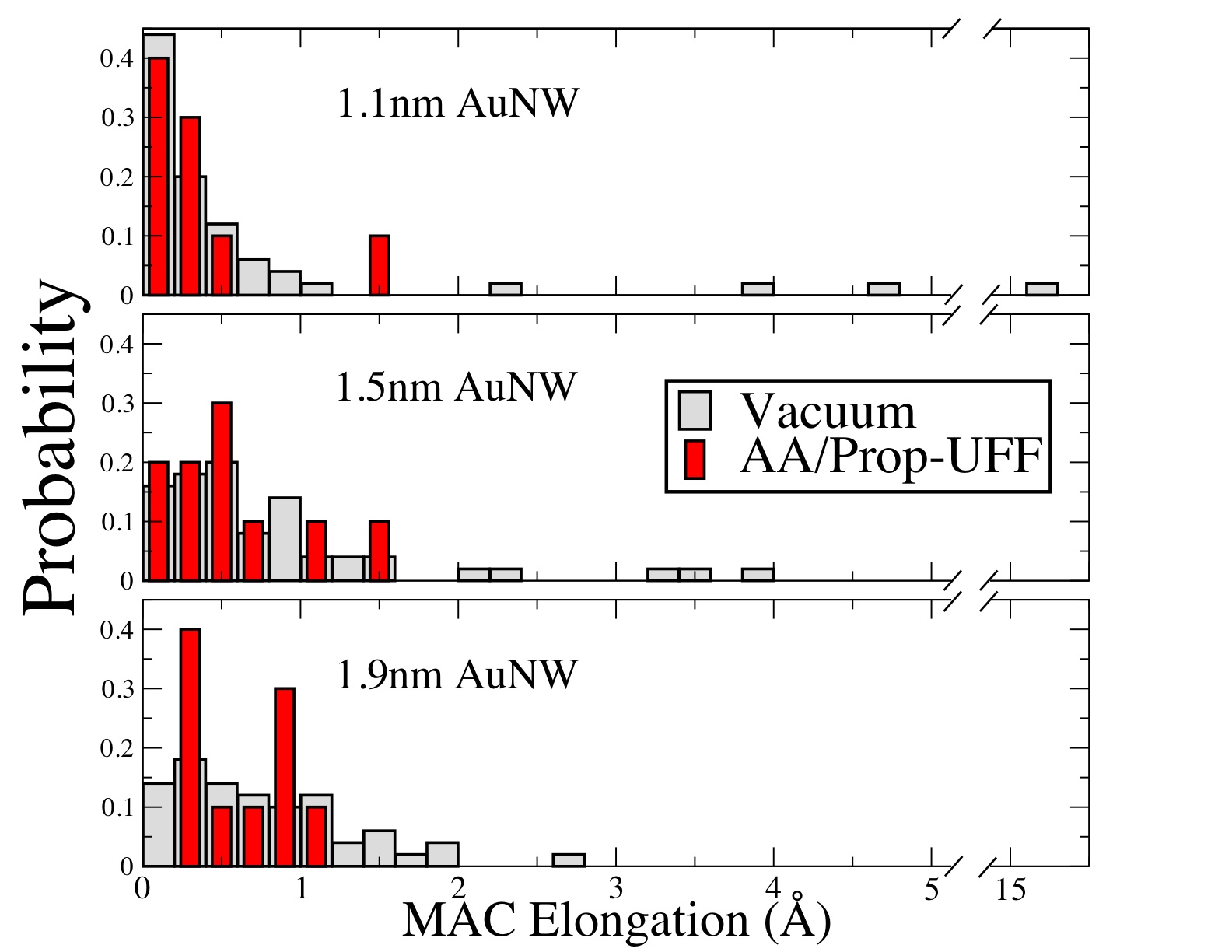}
	\caption{Histograms of monatomic chain stability in vacuum and in AA/Prop-UFF adsorbate for (top) 1.1nm, (middle) 1.5nm, and (bottom) 1.9nm AuNWs. }
	\label{fig:uff_histo}
\end{figure}

\end{document}